\documentclass{aastex61}
\usepackage{hyperref}
\usepackage{amsmath,amstext}
\usepackage[T1]{fontenc}
\usepackage{apjfonts} 
\usepackage[figure,figure*]{hypcap}
\usepackage{booktabs}
\usepackage{multirow}
\usepackage{makecell}
\usepackage{graphicx}
\usepackage{float} 
\usepackage{enumitem}
\usepackage{comment}
\usepackage{color, xcolor}	
\usepackage{diagbox}

\submitjournal{ApJL}

\shorttitle{Modeling Spectroscopic Diversity of TDEs}
\shortauthors{Thomsen et al.}

\newcommand{\helya}{He {\sc ii} Ly\ensuremath{\alpha}\,($\lambda 303$)}
\newcommand{\ha}{H\ensuremath{\alpha}}
\newcommand{\hb}{H\ensuremath{\beta}}
\newcommand{\hd}{H\ensuremath{\delta}}

\newcommand{\nii}{N {\sc ii}}
\newcommand{\niii}{N {\sc iii}}
\newcommand{\niv}{N {\sc iv}}
\newcommand{\nv}{N {\sc v}}

\newcommand{\oiii}{O {\sc iii}}
\newcommand{\oiv}{O {\sc iv}}

\newcommand{\hei}{He {\sc i}}
\newcommand{\heii}{He {\sc ii}}
\newcommand{\heiii}{He {\sc iii}}
\newcommand{\hi}{H {\sc i}}
\newcommand{\hii}{H {\sc ii}}

\newcommand{\civ}{C{\sc iv}}

\newcommand{\niiia}{N {\sc iii}$\lambda 4640$}
\newcommand{\niiib}{N {\sc iii}$\lambda 4100$}

\newcommand{\heiia}{He {\sc ii}$\lambda 4686$}

\newcommand{\oiiif}{O {\sc iii}$\lambda 5592$}

\begin{document}

\title{A Unified Ionization Framework for the Spectroscopic Diversity of Tidal Disruption Events}

\correspondingauthor{Lixin Dai \\
lixindai@hku.hk}

\author[0000-0003-4256-7059]{Lars L. Thomsen}
\affiliation{Department of Physics, The University of Hong Kong, Pokfulam Road, Hong Kong, China}
\affiliation{The Hong Kong Institute for Astronomy and Astrophysics, The University of Hong Kong, Pokfulam Road, Hong Kong, China}

\author[0000-0002-9589-5235]{Lixin Dai}
\affiliation{Department of Physics, The University of Hong Kong, Pokfulam Road, Hong Kong, China}
\affiliation{The Hong Kong Institute for Astronomy and Astrophysics, The University of Hong Kong, Pokfulam Road, Hong Kong, China}

\author[0000-0002-5981-1022]{Daniel Kasen}
\affiliation{Department of Astronomy, University of California, Berkeley, CA 94720, USA}
\affiliation{Department of Physics, University of California, Berkeley, CA 94720, USA}
\affiliation{Nuclear Science Division, Lawrence Berkeley National Laboratory, Berkeley, CA 94720, USA}

\author[0000-0003-2558-3102]{Enrico Ramirez-Ruiz}
\affiliation{Department of Astronomy and Astrophysics, University of California Santa Cruz, 1156 High Street, Santa Cruz, CA 95064, USA}

\author[0000-0002-0326-6715]{Panos Charalampopoulos}
\affiliation{Institute of Space Sciences (ICE-CSIC), Campus UAB, Carrer de Can Magrans, s/n, E-08193 Barcelona, Spain}
\affiliation{Finnish Centre for Astronomy with ESO (FINCA), FI-20014 University of Turku, Finland}

\author[0000-0002-8597-0756]{Giorgos Leloudas}
\affiliation{DTU Space,Department of Space Research and Space Technology, Technical University of Denmark, Elektrovej 327, DK-2800 Kgs. Lyngby, Denmark}

\author[0000-0001-6350-8168]{Brenna Mockler}
\affiliation{Department of Physics and Astronomy, University of California, Davis, CA95616, USA}
\affiliation{The Observatories of the Carnegie Institution for Science, Pasadena, CA91101, USA}

\begin{abstract}
Optical tidal disruption events (TDEs) exhibit extremely broad emission lines ($\approx 10^3$--$10^4~{\rm km~s^{-1}}$) and are observationally classified into four spectroscopic types: H-dominated, He-dominated, H+He, and featureless. The prevalent H+He class often displays Bowen fluorescence lines (notably \niii~and \oiii), features that are rarely observed in active galactic nuclei and whose origin has remained poorly understood. We present the first unified radiative transfer framework that reproduces all four TDE spectroscopic classes using simulations of optically thick, outflowing envelopes with solar composition. Our models successfully capture both the continuum properties and key spectral features, including strong \ha, \heii~and Bowen emissions. We demonstrate that the spectroscopic diversity of TDEs is primarily governed by the gas ionization state, controlled by the ratio of injected luminosity to envelope mass. As the ionization level decreases, the observed sequence of spectroscopic classes emerges naturally, transitioning from featureless to He-dominated, to Bowen-dominated, and finally to H-dominated spectra. We further show that electron scattering in the optically thick outflow is the dominant mechanism responsible for the extreme line widths, linking line profiles directly to the physical properties of the wind. The model also explains the observed correlations with luminosity, black hole mass, and the relative stability of spectral classifications during TDE evolution. This work establishes a unified physical framework for TDE spectroscopy, providing new insight into the emission mechanisms, energetics, and outflow structure of these transient events, and offering a practical pathway for interpreting and fitting observed spectra.
\end{abstract}

\section{Introduction}
\label{sec:intro:lines}
Tidal disruption events (TDEs) arise when a star is torn apart after passing within the tidal radius of a massive black hole (MBH), producing a luminous transient powered by the rapid onset of accretion \citep{Rees88,Evans89,Guillochon13,Rossi21review}. The bound stellar debris returns to the pericenter and dissipates orbital energy through shocks during circularization, leading to the formation of an accretion flow \citep{Guillochon2014,Bonnerot21,Steinberg2024,Xiaoshan2024,Hu2026}. This flow radiates across the electromagnetic spectrum and, in some cases, powers relativistic jets \citep[see e.g.,][]{Giannios2011,DeColle2012,vanVelzen20review,Alexander20review,Saxton21review,Gezari21review}.

A central question in TDE research is the origin of their diverse emission properties. Early theoretical models predicted that the compact accretion disk formed in a TDE would radiate primarily in soft X-rays \citep{Ulmer99}. However, a large fraction of events discovered in optical time-domain surveys, such as the Zwicky Transient Facility (ZTF) and the All-Sky Automated Survey for Supernovae (ASAS-SN), are dominated by near-ultraviolet (NUV) and optical emission and often lack detectable X-ray counterparts. When modeled with a single blackbody, these optical TDEs typically exhibit temperatures of a few $10^4$ K, whereas thermal X-ray TDEs reach temperatures of up to a few $10^5$ K \citep[e.g.,][]{Auchettl2017}. This dichotomy indicates that TDE emission cannot be explained by a standard, unobscured thin accretion disk alone, and instead points to the presence of additional reprocessing layers or complex geometric structures \citep[see reviews by][]{Roth20review,Dai21review}.

One well-studied class of models posits that optically bright TDEs arise from the reprocessing of disk X-ray emission within an optically thick envelope, likely produced by outflows launched during disk formation or subsequent accretion phases \citep{Loeb97,Strubbe09,Ramirez-Ruiz2009,Lodato11,Guillochon2014,Roth16,Lu2020,Bonnerot21,Metzger2022}. Within this framework, \citet{Dai18}, \citet{Thomsen2022}, and \citet{Parkinson25} developed a unified dynamical picture in which super-Eddington accretion drives dense winds that absorb and reprocess high-energy radiation. The efficiency of this reprocessing depends strongly on viewing angle: X-ray photons escape preferentially along the low-density polar funnel, while at higher inclinations they are trapped and reprocessed into optical and ultraviolet emission.

An alternative class of models attributes the optical emission to shocks generated during debris stream collisions, with weak X-ray emission arising from inefficient disk formation \citep{Piran15,Svirski2017}. Although not the focus of this work, this scenario highlights that multiple physical pathways have been proposed to explain the observed diversity of TDE emission.

Comprehensive follow-up observations of optical TDEs have moved well beyond continuum measurements, revealing a rich diversity of spectroscopic features \citep[e.g.,][]{Guillochon2014,Hung19,Leloudas2019,Charalampopoulos2022,Hammerstein2023}. In particular, the discovery of extremely broad and prominent emission lines has led to the identification of several distinct spectroscopic classes:
\begin{itemize}
\item \textbf{TDE-H:} Spectra exhibiting only broad hydrogen Balmer lines (H$\alpha$, H$\beta$, etc.).
\item \textbf{TDE-He:} Spectra dominated by a broad helium~II line (most commonly \heiia).
\item \textbf{TDE-H+He:} Spectra exhibiting both broad hydrogen Balmer and helium~II lines.
\item \textbf{TDE-Featureless:} Spectra lacking prominent broad emission lines.
\end{itemize}
Interestingly, many TDE-H+He events also exhibit a distinctive set of \niii~and \oiii~emission lines associated with Bowen fluorescence \citep{Leloudas2019}. These lines are produced when EUV or X-ray irradiation drives a resonance cascade involving \heii, \oiii, and \niii~ions \citep{Bowen1934}. The detailed physics of this process is discussed in Section~\ref{sec:spectra_model}. In some TDEs (e.g., AT2018dyb), the Bowen \niii~lines can reach fluxes comparable to or even exceeding those of \ha~and \heii.

Systematic studies have revealed clear differences between TDEs that exhibit strong Bowen features and those dominated by hydrogen emission \citep{Charalampopoulos2022,Nicholl2022}. In particular, Bowen TDEs tend to show narrower line widths, with $\mathrm{FWHM} \lesssim 10^4~\mathrm{km~s^{-1}}$, whereas hydrogen-dominated events typically exhibit broader lines with $\mathrm{FWHM} \gtrsim 10^4~\mathrm{km~s^{-1}}$. Another notable distinction is that Bowen fluorescence lines, while prominent in TDEs, are rarely observed in typical active galactic nuclei (AGN). However, recent observations have identified similar features in a small number of flaring or changing-look AGN \citep[e.g.,][]{Trakhtenbrot2019,Frederick2021}.

Understanding the observed diversity of TDEs is essential for uncovering the physical mechanisms that govern their emission. Most theoretical efforts to date have focused on modeling the continuum, including its origin, temperature, and temporal evolution. While these studies have provided important insights, they do not fully exploit the diagnostic power of spectroscopic features. In particular, broad hydrogen Balmer lines, strong \heiia~emission, and Bowen fluorescence features offer direct probes of the density, ionization state, and geometry of the emitting gas, and thus provide a powerful diagnostic of the structure and evolution of the reprocessing environment in TDEs.

Considerable diagnostic power resides in the detailed spectroscopic features of TDEs. Nevertheless, only a limited number of studies have modeled these features, and existing frameworks do not yet capture the full observed diversity. These studies highlight two key factors that shape line formation. First, the chemical composition of the disrupted star can lead to helium- or nitrogen-enhanced abundances \citep{Gezari12,Kochanek2016,Garcia2018,Mockler2022,Miller2023,Mockler2024,Bush2025}, favoring the emergence of He- or N-rich spectra. Second, and more fundamentally, the emergent spectrum is highly sensitive to the gas temperature and ionization state.
Radiative transfer calculations \citep{Roth16,Roth17,Parkinson20,Parkinson22} show that variations in the central radiation field and the properties of the surrounding envelope, including its mass, density, and outflow velocity, can produce a wide range of spectral outcomes. For example, \citet{Roth16} demonstrated that strong ionization suppresses hydrogen Balmer emission, yielding a He-dominated spectrum even for gas with solar composition. Likewise, \citet{Roth17} showed that optically thick envelopes naturally produce very broad line profiles through electron scattering.

Despite these advances, a predictive framework that systematically connects all 
the observed spectroscopic classes to the underlying physical parameters remains lacking.
In this paper, we address this gap by modeling TDE spectroscopic features using state-of-the-art radiative transfer calculations. Our primary goal is to determine whether the observed spectroscopic diversity of TDEs, including the four established classes, can be explained within a single, physically self-consistent framework. To this end, we employ a one-dimensional radiative transfer model of an optically thick, outflowing envelope with solar metallicity, into which central X-ray photons are injected. This setup allows us to isolate the dominant role of photoionization physics, while deferring the effects of intrinsic elemental abundance variations to future work. Through this approach, we aim to constrain the physical conditions of TDE outflows and identify the key parameters that govern their diverse spectroscopic signatures.

This paper is organized as follows. In Section~\ref{sec:method}, we describe our TDE model and the radiative transfer code {\tt SEDONA}. In Section~\ref{sec:result}, we present our simulation results and identify the physical conditions that give rise to each spectroscopic class. In Section~\ref{sec:discussion}, we link our model to observations of TDEs and AGN, and discuss its connection to previous models. Then in Section~\ref{sec:conclusion}, we summarize our findings and outline future work directions.

\section{Methodology}
\label{sec:method}

Our methodology builds on the framework developed in \citet{Roth16}, \citet{Dai18}, and \citet{Thomsen2022}. We represent the system as a spherically symmetric (one-dimensional), optically thick, outflowing envelope surrounding a central X-ray source. Photons injected at the inner boundary propagate through the envelope and are reprocessed via absorption, scattering, and re-emission, transforming the initial high-energy radiation into an emergent spectrum, which is dominated by ultraviolet and optical emission when the envelope is sufficiently optically thick.

The structure of the gas envelope is described in Section~\ref{sec:method:gas}, and the radiative transfer calculations are detailed in Section~\ref{sec:method:sedona}.

\subsection{Set-Up of the Gas Envelope}
\label{sec:method:gas}
In this work, we adopt a simplified, spherically symmetric outflow model to describe the density and velocity structure of the reprocessing gas envelope surrounding a massive black hole (MBH) with mass $M_{\rm BH} = 10^6\,M_\odot$. The envelope extends from an inner radius $R_{\rm in}$ to an outer radius $R_{\rm out}$. We assume the gas density follows a power-law profile,
\begin{equation} \label{eq:density}
\rho = \rho_0 \ (r/R_g)^{-q},
\end{equation}
where $R_g \equiv GM_{\rm BH}/c^2$ is the gravitational radius.

For a given total envelope mass $M_{\rm env}$, the density normalization constant $\rho_0$ is determined by integrating the density profile over the envelope and equating the result to $M_{\rm env}$:
\begin{equation} \label{eq:mass}
    M_{\rm env} = \int_{R_{\rm in}}^{R_{\rm out}} 4\pi \rho_0 \left( \frac{r}{R_{\rm g}} \right)^{-q} r^2 \, dr
                = 4\pi \rho_0 R_{\rm g}^{\,q} \left[ \frac{r^{3-q}}{3-q} \right]_{R_{\rm in}}^{R_{\rm out}}, \qquad (q < 3).
\end{equation}

For the envelope velocity profile, we adopt a simplified outflow model that captures the essential physics of disk-driven winds. Such winds are expected to undergo an initial acceleration phase before reaching a terminal velocity or, in some cases, transitioning to a decelerating flow \citep{Dai21review,Bu2023,Yang2024}. In addition, they are likely to be collimated by magnetic or radiation pressure.

\begin{figure}
    \centering
    \includegraphics[width=0.7\linewidth]{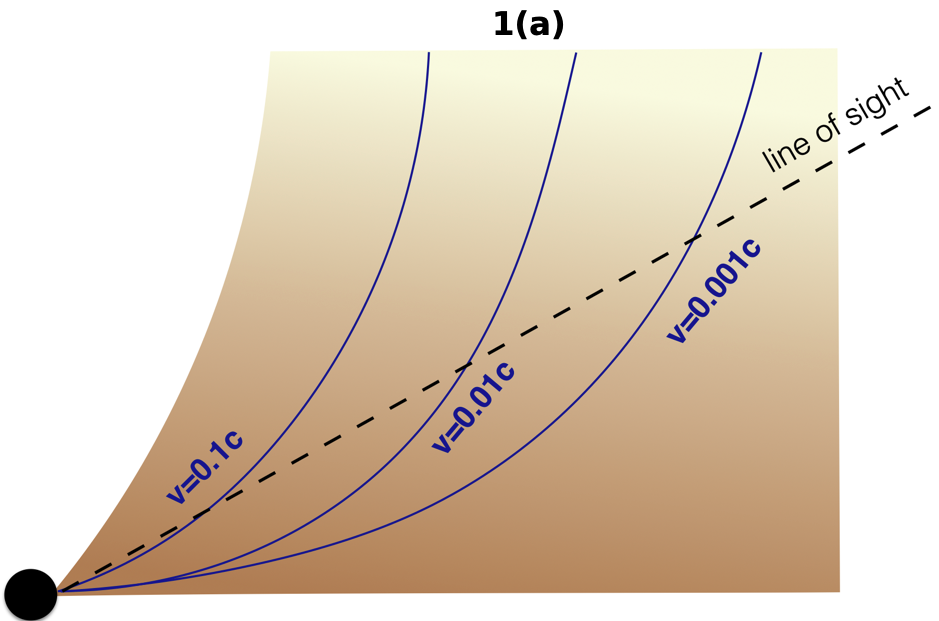}
    \includegraphics[width=0.7\linewidth]{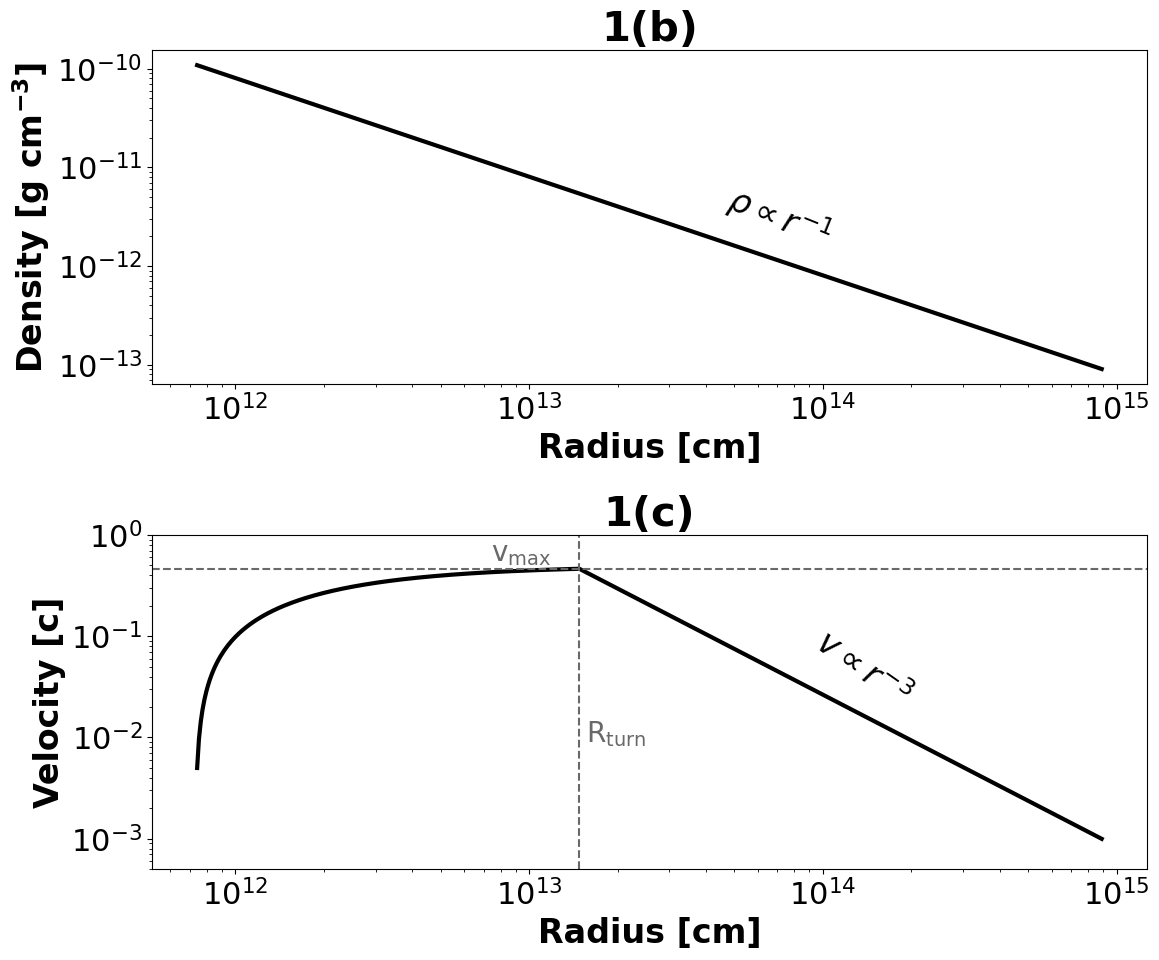}
    \caption{Physical setup and model profiles for the reprocessing envelope. Panel (a) illustrates a schematic two-dimensional velocity structure motivated by simulations of super-Eddington accretion flows. Radiation and magnetic pressure collimate the outflow, causing surfaces of constant radial velocity to bend toward the polar axis. As a result, a photon traveling along a fixed radial line of sight intersects regions at progressively larger cylindrical radii, effectively sampling a velocity profile that increases, reaches a maximum, and then decreases. Panels (b) and (c) show the fiducial density and velocity profiles adopted in our radiative transfer simulations. The density profile corresponds to a total envelope mass of $M_{\rm env} = 0.2\,M_\odot$, consistent with the mass expected to have been processed through the system shortly after the peak fallback rate of a solar-mass star. In panel (c), we highlight key features of the velocity profile, including the turnover radius $R_{\rm turn}$, where the flow transitions from acceleration to deceleration, and the corresponding maximum velocity $v_{\rm max}$.}
    \label{Fig:velocity}
\end{figure}

As a result, surfaces of constant radial velocity are not purely radial but instead curve outward, as illustrated in Figure~\ref{Fig:velocity}(a). This geometry introduces a useful simplification: along a given radial line of sight, the flow can be approximated as sampling winds launched at progressively larger radii, corresponding to an effective deceleration of the velocity profile along that ray.

Our test simulations indicate that such acceleration--deceleration profiles produce stronger optical emission lines than purely monotonic velocity structures. While the detailed origin of this behavior remains to be explored, this result also provides practical motivation for adopting the velocity profile used in this work.

To capture this line-of-sight behavior, we adopt a composite velocity profile consisting of two components: (1) an inner acceleration zone, in which the wind speed increases following functional forms similar to those in \citet{Shlosman1993,Kara16,Parkinson20}; and (2) an outer deceleration zone, in which the velocity declines as a power law with radius. 
The resulting velocity profile is given by:
\begin{equation} \label{eq:vel}
v(r) =
\begin{cases}
v_0 + (v_t - v_0) \times \frac{r - R_{\rm in}}{r + R_{\rm acc}}, & r \le R_{\rm turn}, \\[8pt]
v_{\rm max} \left( \dfrac{r}{R_{\rm in}} \right)^{-p}, & r \ge R_{\rm turn}.
\end{cases}
\end{equation}
Here, $v_0$ denotes the wind launch velocity at the inner boundary $R_{\rm in}$, while $v_t$ represents the terminal velocity the flow would approach in the absence of deceleration. The parameter $R_{\rm acc}$ sets the characteristic scale of the acceleration region. The velocity reaches a maximum value $v_{\rm max} = v(R_{\rm turn})$ at the turnover radius $R_{\rm turn}$, where the flow transitions from acceleration to deceleration.

We conducted a suite of test simulations to explore the parameter space of our toy model, including the power-law indices $p$ and $q$, the asymptotic terminal velocity $v_t$, and other key variables. Based on this exploration, we adopt the following fiducial parameter set:
$R_{\rm in} = 5 R_g$, $R_{\rm out} = 6000 R_g$, $R_{\rm acc} = 20 R_g$, $R_{\rm turn} = 100 R_g$, $v_0 = 0.005c$, $v_t = 0.5c$, $v_{\rm max} = 0.46c$, $q = 1$, and $p = 3$. These values are chosen to be broadly consistent with expectations from theoretical models and observational constraints on TDE outflows. In addition, this configuration robustly produces strong optical emission lines while remaining computationally tractable. The corresponding density and velocity profiles for this fiducial model are shown in panels (b) and (c) of Figure~\ref{Fig:velocity}, respectively.

A notable feature of our model is the adoption of a shallow density profile with a power-law index $q = 1$, which enables the efficient production of optical emission lines. This choice reduces the overall optical depth while maintaining relatively high densities in the outer envelope, creating favorable conditions for line formation. Moreover, such a shallow profile is physically well motivated. It is a robust prediction of winds launched from hot accretion flows, including Advection-Dominated Accretion Flows (ADAF) \citep{Yuan12}, super-Eddington accretion \citep{Dai18}, and ZEro-BeRnoulli Accretion (ZEBRA) solutions \citep{Coughlin2014,Wu2018}. These models consistently yield density profiles shallower than the canonical $q = 2$ scaling expected for a steady, spherical, constant-velocity wind.

As a consequence, the wind mass outflow rate, $\dot{M}_{\rm wind} = \rho v r^2$, is not constant with radius in our model. This behavior reflects the anisotropic nature of realistic disk winds, where geometric collimation leads to variations in mass flux across streamlines, in contrast to the idealized case of a spherically symmetric, conserved flow.

The gas envelope is initialized with a temperature profile of $T = 2 \times 10^6~\mathrm{K}\,(r/R_{\rm in})^{-0.5}$. During the radiative transfer calculations, the temperature is iteratively updated to achieve self-consistency with the radiation field. The initial profile is set slightly hotter than the expected equilibrium solution to accelerate convergence during the early stages of the simulation.

Building on this fiducial setup, we perform a series of radiative transfer simulations to map the resulting spectra across the two-dimensional parameter space defined by the total envelope mass ($M_{\rm env}$) and the injected X-ray luminosity ($L_{\rm inj}$). The detailed properties of the injected photon distribution are described in the following section.

\subsection{Radiative Transfer Simulations}
\label{sec:method:sedona}

We use the Monte Carlo radiative transfer code {\tt SEDONA} \citep{Kasen06} to model the interaction between radiation and the gas in the reprocessing envelope. As our focus is on the role of ionization in shaping the emergent spectra, we adopt solar abundances throughout.

In contrast to previous studies that included a limited set of elements, typically H, He, and O \citep{Roth16,Dai18,Thomsen2022}, our model incorporates a more comprehensive composition. Specifically, we include hydrogen (H), helium (He), carbon (C), nitrogen (N), oxygen (O), sodium (Na), magnesium (Mg), silicon (Si), sulfur (S), calcium (Ca), titanium (Ti), and iron (Fe). This expanded elemental set enables a more accurate treatment of radiative processes and, importantly, allows us to capture key spectral features such as Bowen fluorescence lines, which depend sensitively on nitrogen, oxygen, and helium.

The {\tt SEDONA} code includes the key radiative processes relevant to our calculations, namely electron scattering, free-free, bound-bound, and bound-free interactions. Bound-bound transition data are drawn from the {\tt CMFGEN} atomic database. To maintain computational efficiency, we include a subset of the most important transitions for elements beyond hydrogen and helium. For iron, we track ionization states from Fe II to Fe VII.

Line interactions from bound-bound and bound-free processes are treated using an effective broadening velocity of $10^7 \ {\rm cm \ s^{-1}} \ (10^2 \ {\rm km \ s^{-1}})$, which ensures adequate resolution within the Monte Carlo framework (see Appendix \ref{sec:method:line} for further details).
The bound-free cross sections are calculated using the hydrogenic approximation:
\begin{equation}
\sigma_{\rm bf}(\nu, n_{\rm pq}, Z) =
\begin{cases}
\sigma_0 \ n_{\rm pq} \ Z^{-2} \ (\nu / \nu_{\rm thresh})^{-3}, & \nu > \nu_{\rm thresh}, \\
0, & \text{otherwise},
\end{cases}
\end{equation}
where $\sigma_0 = 6.3 \times 10^{-18}  \text{cm}^2$ is a scaling constant, $n_{\rm pq}$ is the principal quantum number, $Z$ is the effective nuclear charge of the ion, $\nu$ is the photon frequency, and $\nu_{\rm thresh}$ is the ionization threshold frequency corresponding to the binding energy of the least bound electron.

Our implementation includes two additional physical treatments. First, we incorporate electron Comptonization following \citet{Roth17}. Second, the ionization states and level populations of bound electrons are computed under non-local thermal equilibrium (nLTE) conditions using the methodology of \citet{Roth16}.

We initialize the radiative transfer calculation by injecting blackbody radiation with a temperature of $T = 10^6\,\mathrm{K}$ at the inner boundary. This temperature is characteristic of the inner accretion disk around a $10^6\,M_\odot$ MBH. We adopt a constant injection temperature across all simulations, which is a reasonable approximation given that the inner radius is fixed. Although the luminosity varies by up to a factor of 32, the corresponding change in temperature scales as $T \propto L^{1/4}$ and is therefore limited to a factor of $\sim 2.4$. As we show in the next section, most configurations efficiently absorb and reprocess the injected X-rays to lower energies, so variations in the injection temperature do not significantly affect our main results.

Injected photons are propagated through the spherically symmetric gas envelope (described in Section~\ref{sec:method:gas}) using a three-dimensional Monte Carlo scheme. Photon packets are tracked until they either escape beyond the outer boundary at $R_{\rm out} = 6000\,R_g$ or are absorbed, transferring their energy to the gas.

The simulations are performed iteratively. In each iteration, we update the local gas temperature, ionization states, and level populations based on the cumulative interaction history of photon packets within each radial zone. These updates are carried out under the assumption of statistical equilibrium and repeated until a steady-state solution is reached. Although convergence is typically achieved within 15--30 iterations, we run each simulation for at least $n \geq 40$ iterations to ensure stability.

\section{Results}
\label{sec:result}
Using the spherically symmetric outflow setup described in Section~\ref{sec:method:gas}, we perform an extensive grid of radiative transfer simulations with {\tt SEDONA} (Section~\ref{sec:method:sedona}). We systematically vary the two primary physical parameters: the total gas envelope mass, \( M_{\rm env} \), and the central injected X-ray luminosity, \( L_{\rm inj} \). The values explored in this parameter space are:
\[
\begin{aligned}
M_{\rm env} &= (0.05,\ 0.1,\ 0.15,\ 0.2,\ 0.25) \, M_\odot, \\
L_{\rm inj} &= (0.5,\ 1,\ 2,\ 4,\ 8,\ 16) \times 10^{44} \, \text{erg s}^{-1}.
\end{aligned}
\]
When varying $L_{\rm inj}$, we keep the injected radiation temperature fixed at $T = 10^6$~K and adjust only the radiation flux. This setup allows us to isolate the role of luminosity in setting the ionization state of the gas.

This parameter study maps the resulting spectroscopic features, including the continuum level, line strengths and types, and ionization structure, across the space defined by these two fundamental drivers. In this section, we present the simulated spectra, identify the conditions that give rise to the four established TDE spectroscopic classes, and show that their emergence is primarily governed by the gas ionization state.

\subsection{Envelope Mass and Radiation Luminosity Dependence of Reprocessed TDE Spectra}
\label{sec:3.1}

\begin{figure}
    \centering
    \includegraphics[width=0.5\linewidth] {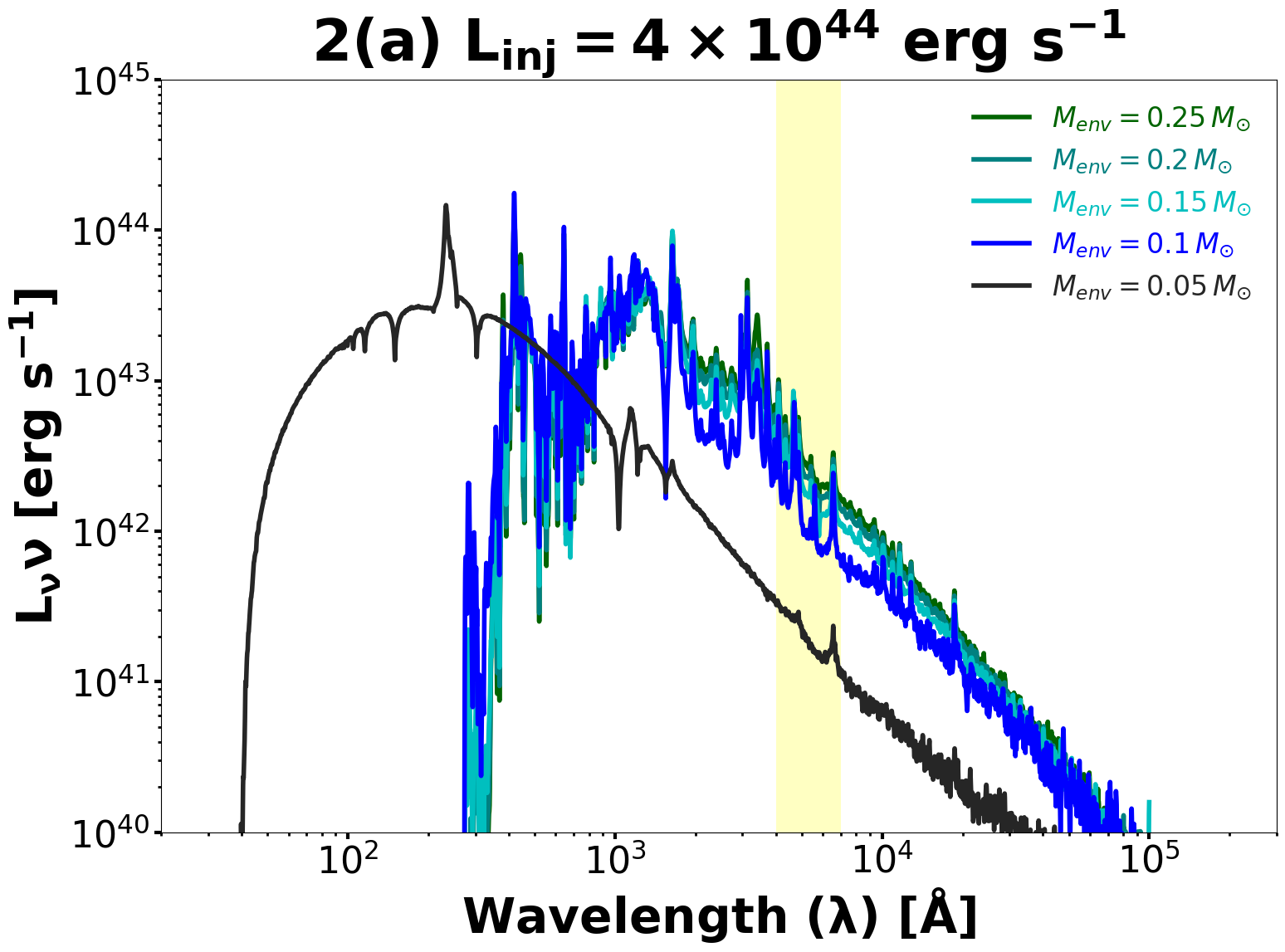}\hfill
    \includegraphics[width=0.5\linewidth]{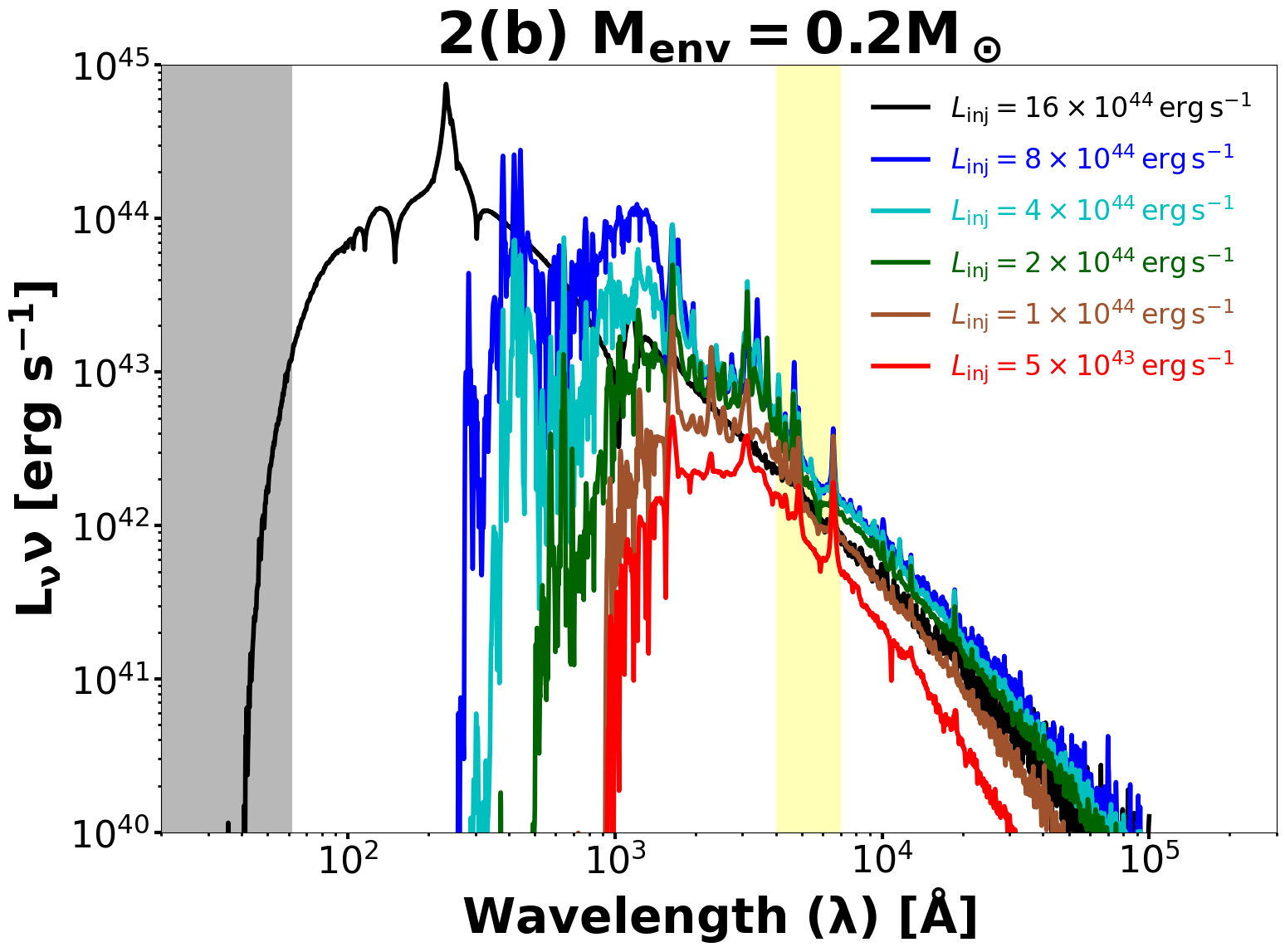} \hfill
    \includegraphics[width=0.8\linewidth]{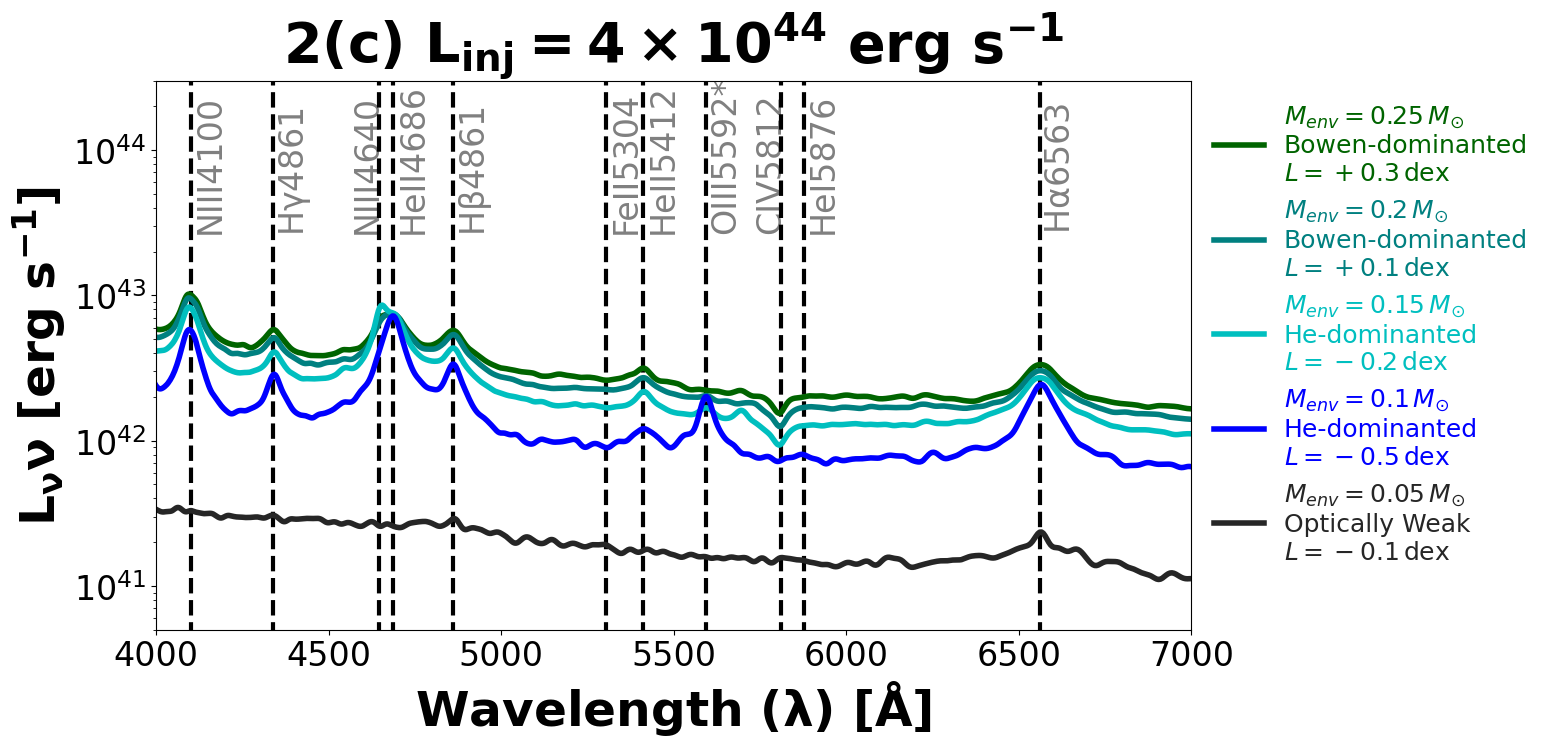} 
    \includegraphics[width=0.8\linewidth]{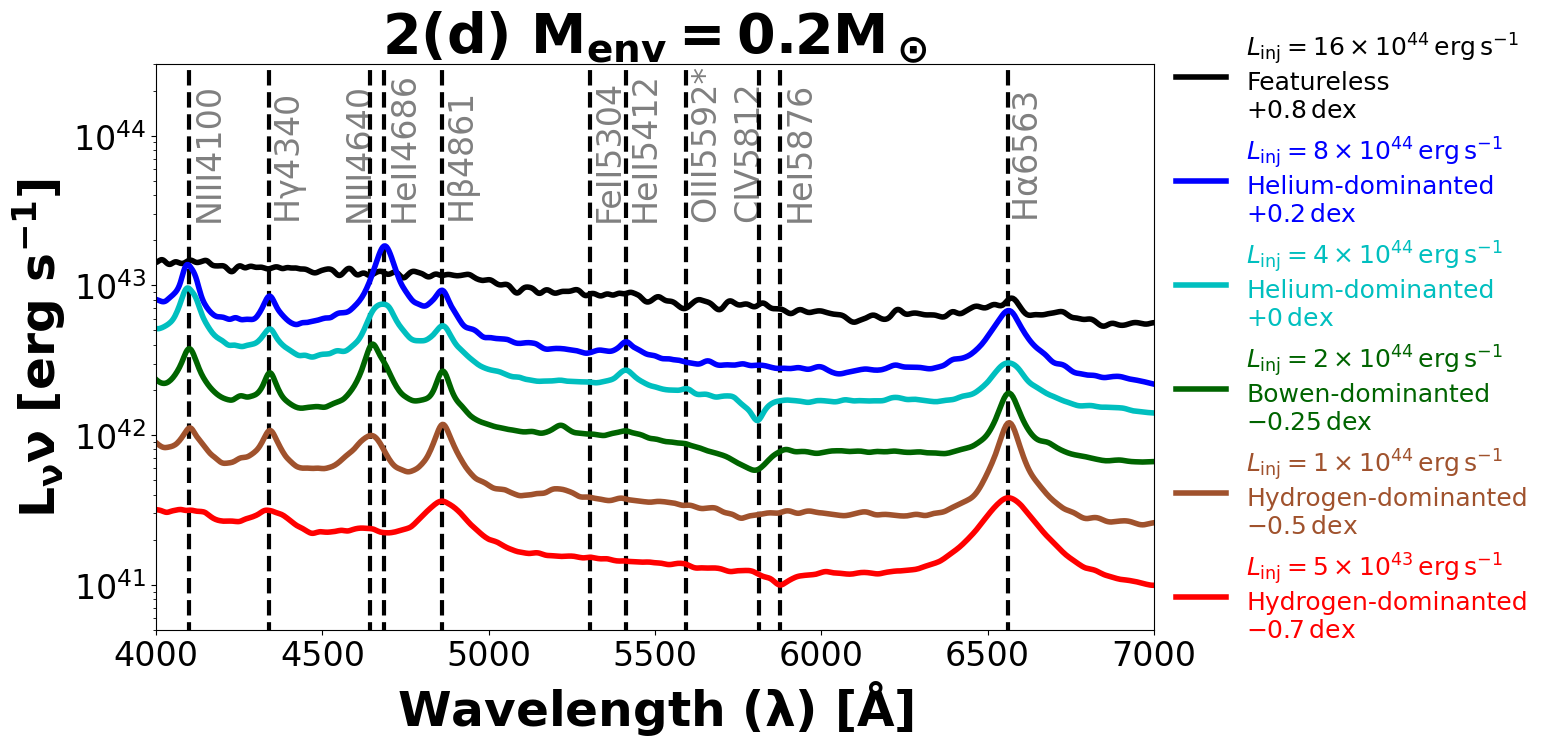}
    \caption{Reprocessed TDE spectra as a function of envelope mass and injected luminosity. Panels (a) and (c) show the effect of varying the injected luminosity $L_{\rm inj}$ at fixed envelope mass, while panels (b) and (d) show the effect of varying the envelope mass $M_{\rm env}$ at fixed luminosity. The top row (a, b) presents the full reprocessed spectra, and the bottom row (c, d) shows a zoomed-in view of the optical wavelength region (highlighted in yellow). Spectra are color-coded by their dominant emission feature: red for hydrogen-dominated, blue for helium-dominated, green for Bowen-dominated, and black for featureless.
    }
    \label{fig:spectra}
\end{figure}

From the simulation grid, we find that, as expected from photoionization physics, the ionization state and resulting spectroscopic features exhibit a clear and systematic dependence on both the envelope mass \( M_{\rm env} \) and the injected X-ray luminosity \( L_{\rm inj} \). Figure~\ref{fig:spectra} summarizes the characteristic reprocessed spectra produced across this parameter space.

We first examine the effect of envelope mass by fixing the injected luminosity at $L_{\rm inj} = 4 \times 10^{44} \ \rm erg \ s^{-1}$ and varying $M_{\rm env}$. The resulting spectra are shown in Figure~\ref{fig:spectra} panel (a), with a zoomed-in view of the optical band presented in panel (c). 

We next isolate the influence of luminosity by fixing the envelope mass at $M_{\rm env} = 0.2\,M_\odot$ and varying $L_{\rm inj}$. The corresponding full and optical-band spectra are shown in panels (b) and (d) of Figure~\ref{fig:spectra}, respectively.

The effects of the two primary parameters on the emergent emission can be summarized as follows. Increasing \( M_{\rm env} \) provides more reprocessing material, which lowers the equilibrium temperature of the gas and reduces its ionization level. Conversely, increasing \( L_{\rm inj} \) raises the gas temperature and ionization level, thereby decreasing the efficiency of X-ray reprocessing. As a result, both the continuum and line properties of the spectra vary systematically across this parameter space:
\begin{enumerate}
    \item {Optical continuum versus envelope mass:} Figure~\ref{fig:spectra}(a) shows that as $M_{\rm env}$ increases, the optical continuum brightens slightly. This behavior reflects the increased optical depth from higher gas density and lower temperature, which enhances the reprocessing of X-ray photons into lower-energy emission. The overall spectrum becomes relatively insensitive to $M_{\rm env}$ once the envelope is sufficiently massive to achieve near-complete reprocessing. This trend is consistent with earlier one-dimensional models \citep{Roth16} and viewing-angle dependent disk models \citep{Dai18,Thomsen2022,Parkinson25}.
    
    \item {Optical continuum versus injected luminosity:} Figure~\ref{fig:spectra}(b) shows that while the total escaped bolometric luminosity increases with $L_{\rm inj}$, the optical luminosity exhibits a non-linear response. At the highest luminosities, the optical continuum increases only modestly, while a significant fraction of X-ray photons escapes. This behavior arises because the intense radiation field drives the gas toward a highly ionized state, reducing bound-free and bound-bound absorption and thus limiting reprocessing. The resulting spectrum becomes nearly featureless, reflecting this highly ionized regime.
    
    \item {Optical spectral lines versus envelope mass:} The evolution of optical spectral features with $M_{\rm env}$ is shown in Figure~\ref{fig:spectra}(c). At low $M_{\rm env}$, the spectrum exhibits only very weak, broad hydrogen Balmer lines, which may be undetectable in observations; such events could be categorized as featureless. Furthermore, the minimal reprocessing at very low $M_{\rm env}$ makes these events difficult to detect with optical transient surveys. Hereafter, we categorize the simulated TDE spectra as optically weak if $L_{5100}<5\times 10^{41} {\rm erg~s^{-1}}$, using this threshold as a rough guideline.  As $M_{\rm env}$ increases, strong broad H Balmer and \heiia~lines emerge. The \heiia~line dominates at lower $M_{\rm env}$, while H$\alpha$ becomes dominant at higher $M_{\rm env}$. Bowen (N~III/O~III) lines also appear and reach peak strength at intermediate values of $M_{\rm env}$.
    
    \item {Optical spectral lines versus injected luminosity:} Figure~\ref{fig:spectra}(d) shows a clear progression of dominant spectral features with increasing $L_{\rm inj}$. The spectra evolve from H$\alpha$-dominated to Bowen (\niii) dominated, then to \heiia-dominated, and finally to a featureless state at the highest luminosities. This trend is qualitatively similar to that seen with decreasing $M_{\rm env}$. However, we note that the featureless events driven by very high $L_{\rm inj}$ also have high optical luminosities and are distinct from those associated with very low $M_{\rm env}$, which suffer from minimal reprocessing and therefore are optoically weak.
\end{enumerate}

One can see that increasing $M_{\rm env}$ and decreasing $L_{\rm inj}$ produce similar spectroscopic effects. Therefore, we propose that the gas ionization state controlled by these two parameters is the ultimate parameter determining the spectroscopic features observed in TDEs. At high ionization (high $L_{\rm inj}$ or low $M_{\rm env}$), the spectrum is nearly featureless. As the ionization decreases, broad \heii~and hydrogen Balmer lines emerge, with \heii~initially dominant. At intermediate ionization, Bowen \niii~emission appears and can become the strongest feature. At the lowest ionization states, \ha~dominates while other lines weaken. A detailed analysis of these trends and their physical origin is presented in Sections~\ref{sec:spectra_trends} and \ref{sec:spectra_model}.

To first order, ionization in the outer envelope's line-emitting region depends primarily on density and injected luminosity. We therefore define a modified ionization parameter, $\xi_{\text{mod}} \propto L_{\text{inj}} / \rho \propto L_{\text{inj}} / M_{\text{env}}$, to explain how the ionization level depends on the envelope mass and injected luminosity, though we caution that this does not represent the physical ionization parameter. Because optical emission lines originate in optically thick outer regions, the local ionizing flux differs significantly from $L_{\text{inj}}$. Even so, tracking the competing effects of $L_{\text{inj}}$ and $M_{\text{env}}$ offers a valuable heuristic for organizing the observed spectroscopic trends.

Although our focus is on the dominant emission features, the models also produce weaker lines from elements such as carbon, oxygen, and iron. We caution that the prominent \oiiif~feature in Figure~\ref{fig:spectra} is likely not physical. It arises from the limited set of bound--bound transitions included for computational efficiency and disappears in test simulations using a more complete atomic dataset (see Appendix~\ref{App:atom}).

\subsection{Properties of the Modeled Optical Emission Lines in TDEs} \label{sec:spectra_trends}

In this section, we analyze the fluxes and widths of the primary optical emission lines produced in our simulations: \ha, \hb, \heiia, and \niiia. We exclude the \niiib~line from this analysis due to significant blending with \hd, although strong \niiib~emission is present in many of our models.
Prior to measuring the line properties, we subtract the underlying continuum, modeled as two power laws fitted to the simulated optical spectrum. The following wavelength regions are excluded from the continuum fit: \ha~(6000--7000~{\AA}) and the \hb+ \heiia+ \niiia~complex (4500--5300~{\AA}).

For line fitting, we note that although including both broad and narrow components can improve the statistical fit (i.e., reduce $\chi^2$), the resulting multi-component structure complicates direct comparison with observational. To maintain a consistent and tractable framework, we follow the approach of \citet{Charalampopoulos2022} and fit each emission line with a single Gaussian or Lorentzian profile, selecting the functional form that minimizes $\chi^2$.

\subsubsection{Classification of the Dominant TDE Emission Line Types}
\begin{figure}
    \centering
    \includegraphics[width=0.8\linewidth]{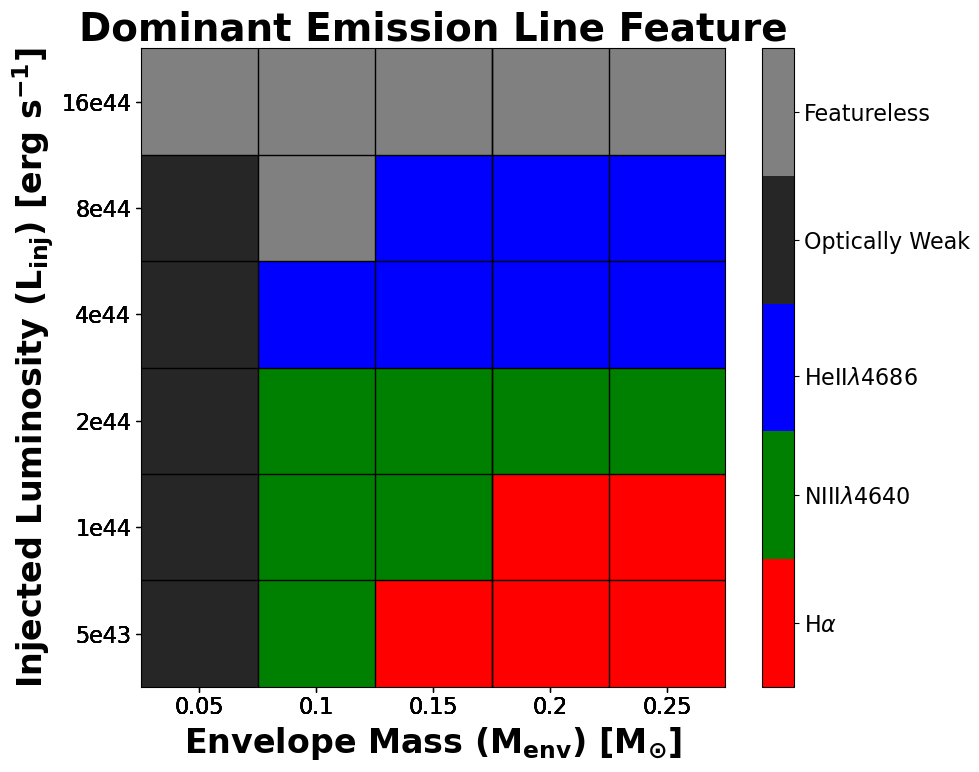}
    \caption{Dominant optical emission line type across the TDE parameter space. For each simulation in the grid, we classify the dominant optical emission feature based on the maximum continuum-subtracted flux among the \ha, \heiia, and \niiia~lines. A smooth transition in the dominant line type emerges as the effective ionization parameter, set by the combined influence of $L_{\rm inj}$ and $M_{\rm env}$, varies across the parameter space.}
    \label{fig:TDE_Phase}
\end{figure}
After fitting \ha, \hb, \heiia, and \niiia, we identify the line with the largest flux in each model. The resulting dominant optical line type across the $L_{\rm inj}$--$M_{\rm env}$ parameter space is shown in Figure~\ref{fig:TDE_Phase}. A clear monotonic trend emerges: the dominant feature transitions from featureless to \heiia, to Bowen-dominated (\niii), and finally to \ha~when moving from the high-ionization regime (top-left; highest $L_{\rm inj}$ and lowest $M_{\rm env}$) to the low-ionization regime (bottom-right; lowest $L_{\rm inj}$ and highest $M_{\rm env}$).

This sequence is consistent with the physical picture presented in Section~\ref{sec:3.1}, supporting the interpretation that the optical emission line strengths in TDEs are primarily governed by the gas ionization state.

In our simulation grid, \heiia-dominated TDEs occur for $L_{\rm inj} \gtrsim 2 \times 10^{44}~\text{erg s}^{-1}$, corresponding to several times the Eddington luminosity for a $10^6\,M_\odot$ black hole. Bowen (\niii)-dominated spectra emerge when $L_{\rm inj} \sim L_{\rm Edd}$, while H$\alpha$ becomes dominant at sub-Eddington luminosities. We emphasize that these thresholds are derived for envelope masses in the range $M_{\rm env} = 0.1$--$0.25\,M_\odot$ within our one-dimensional framework. Variations in the envelope mass or structural properties (e.g., velocity profile, density slope, or size) can shift the critical luminosity associated with each spectral class.

Because the dominant line type is set by the ionization state, simultaneous changes in $L_{\rm inj}$ and $M_{\rm env}$ that preserve their ratio are expected to leave the spectral classification largely unchanged. We return to the implications for TDE spectral evolution in Section~\ref{sec:evolution}.

Here, we emphasize the distinction between the classification scheme derived from our physical model and the observational taxonomy used in the literature. Observationally, optical TDEs are grouped into four spectroscopic classes: TDE-H, TDE-He, TDE-H+He, and TDE-featureless. The TDE-H+He class, which often exhibits strong Bowen \niii~emission \citep{Leloudas2019,Velzen2021}, is sometimes further identified as a TDE-Bowen subclass. 

These observational categories do not map one-to-one onto the dominant line types identified in our simulations (Figure~\ref{fig:TDE_Phase}). While our models provide high spectral resolution and well-defined line strengths, real observations are subject to physical and instrumental limitations. For example, a TDE with a strong \heiia~line and a weak \ha~line may be classified observationally as either TDE-He or TDE-H+He, depending on whether the Balmer emission is detectable above the continuum. In our framework, such a case is unambiguously classified as He-dominated.

To bridge this gap and provide more practical diagnostics, we quantify the relative line strengths by computing flux ratios between \ha~and other prominent lines (\heiia, \niiia), thereby establishing comparative baselines. For the remainder of this work, however, we adopt the dominant-line classification defined by our simulations to maintain internal consistency and a clear physical interpretation.

\subsubsection{Emission Line Flux Ratios}
\begin{figure}
    \centering
    \includegraphics[width=0.45\linewidth]{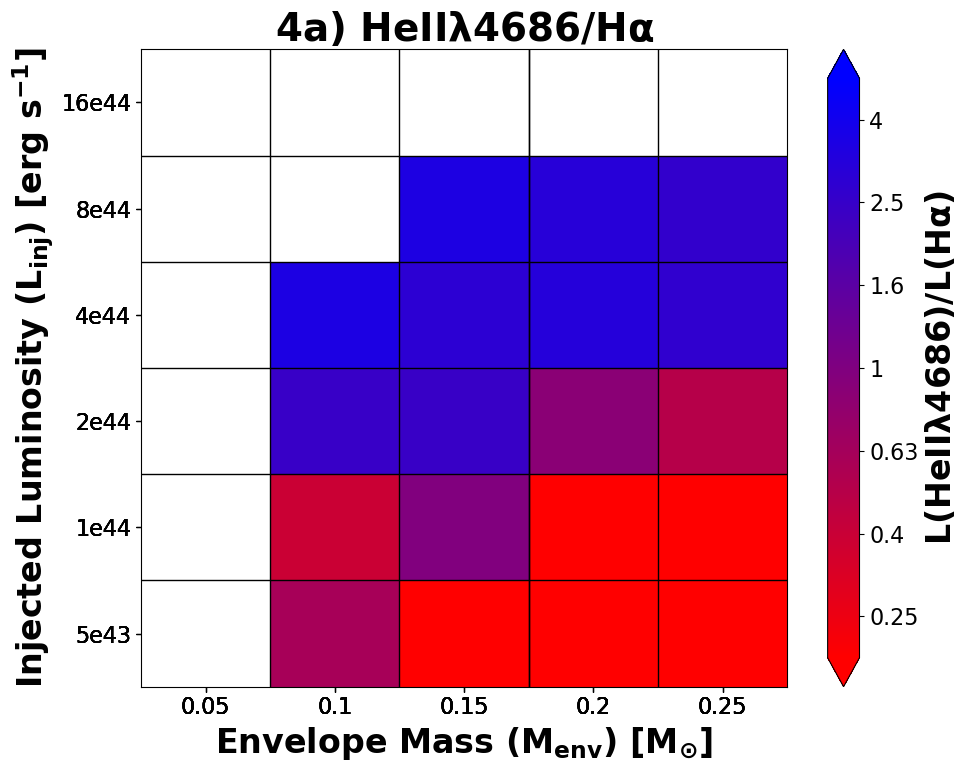}
    \includegraphics[width=0.45\linewidth]{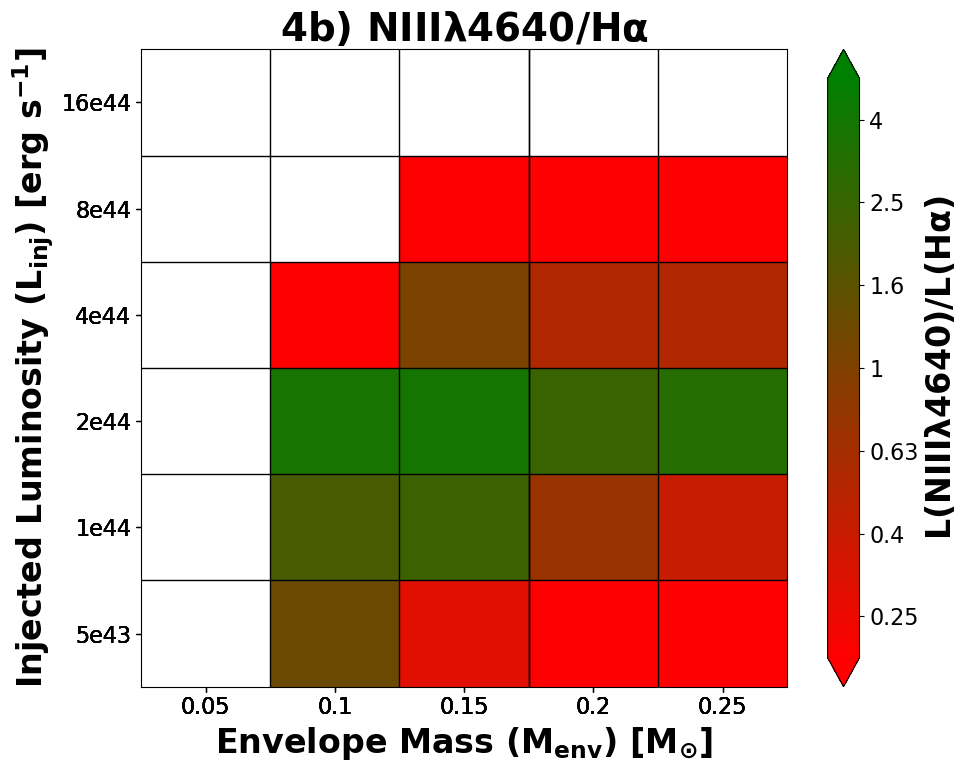}
    \includegraphics[width=0.45\linewidth]{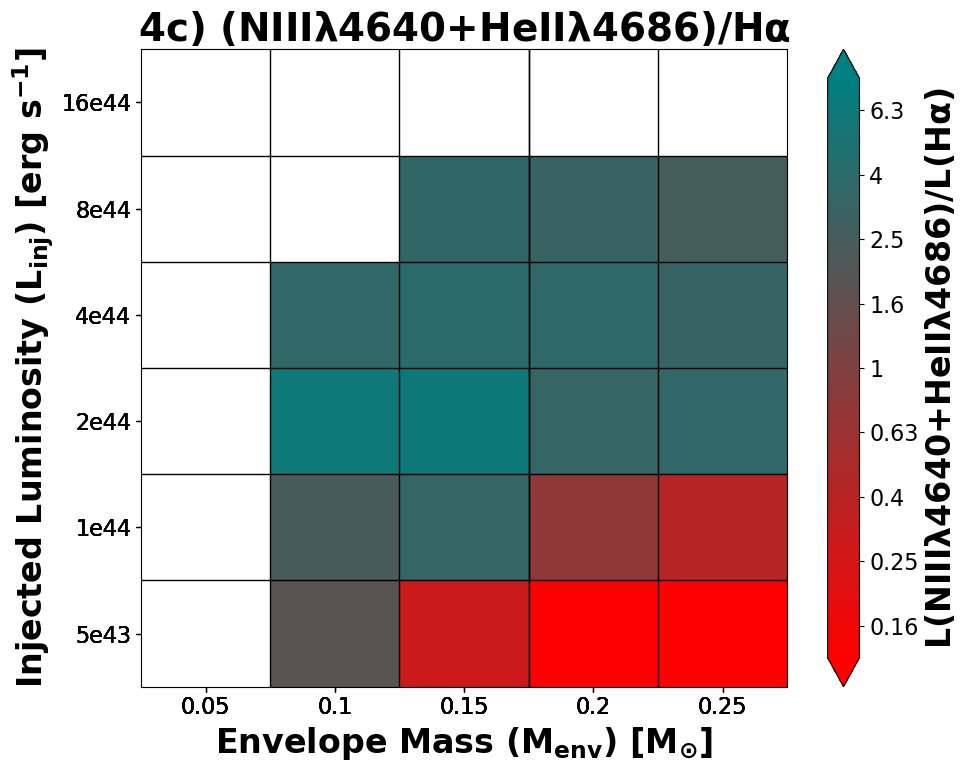}    
    \includegraphics[width=0.45\linewidth]{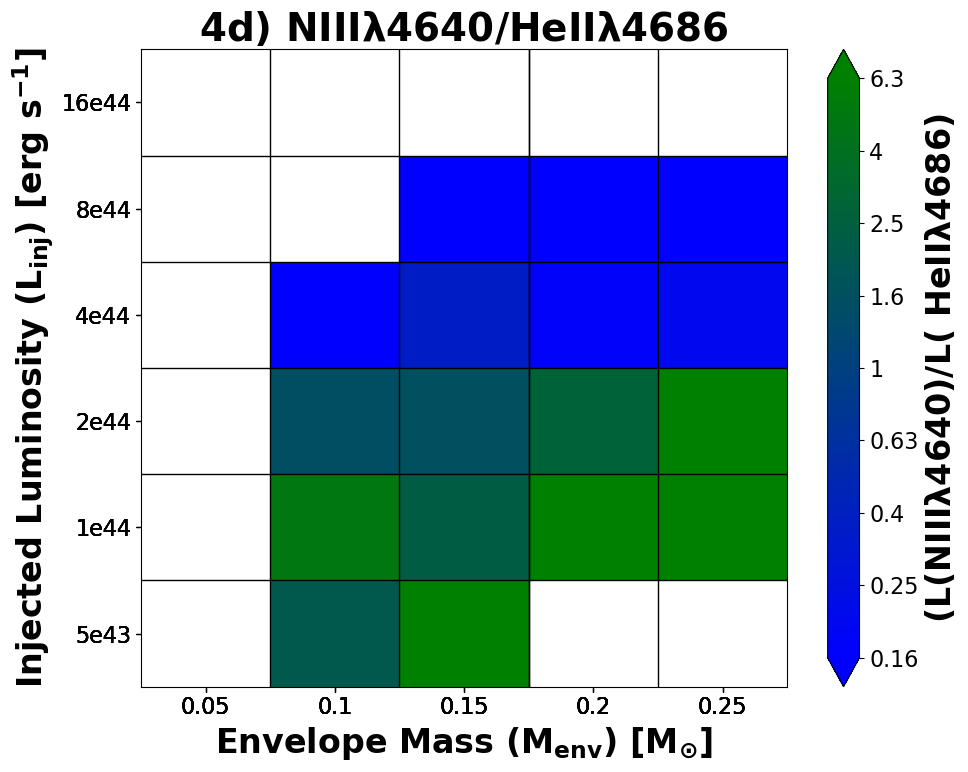}    
    \caption{Diagnostic line flux ratios as a function of ionization state. 
(a) \heiia/\ha: The \heiia/\ha~ratio increases with ionization (higher $L_{\rm inj}$ or lower $M_{\rm env}$), with He~II clearly dominating in the high-ionization regime. 
(b) \niiia/\ha: \niiia~exceeds \ha~only within a narrow region of parameter space corresponding to the ionization conditions optimal for Bowen fluorescence. 
(c) (\niiia~+ \heiia)/\ha: The combined flux of \niiia~and \heiia~relative to \ha~becomes dominant at moderate to high ionization, peaking where both lines are present. 
(d) \niiia/\heiia: This panel highlights the anti-correlation between \heiia~and \niiia~emission, with \heiia~dominating at high ionization and \niiia~at lower ionization.
    }
    \label{fig:line_ratios}
\end{figure}
To quantify the relationship between different emission lines, we compute several key flux ratios from our simulated spectra, shown in Figure~\ref{fig:line_ratios}. Our main findings are as follows:

\begin{itemize}
    \item \heiia~/~\ha: As shown in Figure~\ref{fig:line_ratios}(a), the \heiia/\ha~ratio increases monotonically with the ionization parameter (i.e., with increasing $L_{\rm inj}$ or decreasing $M_{\rm env}$). In Appendix~\ref{sec:appendix_He}, we demonstrate how this ratio is further influenced by the envelope compactness, consistent with the results of \citet{Roth16}.
    
    \item \niiia~/~\ha: In contrast, the \niiia/\ha~ratio (Figure~\ref{fig:line_ratios}(b)) exhibits a non-monotonic dependence on ionization. The \niiia~line exceeds \ha~only within a narrow range of moderate ionization, centered around $L_{\rm inj} = (1$--$2) \times 10^{44}~{\rm erg~s^{-1}}$ for our fiducial $M_{\rm env}$ and $M_{\rm BH}$. This corresponds to the ionization conditions optimal for Bowen fluorescence.
    
    \item (\niiia~+ \heiia)~/~\ha: Because \niiia~and \heiia~are significantly blended, we also consider their combined flux relative to \ha. As shown in Figure~\ref{fig:line_ratios}(c), this ratio exhibits a more monotonic trend with ionization and exceeds unity at moderate to high ionization, where both lines contribute significantly.
    
    \item \niiia~/~\heiia: Finally, Figure~\ref{fig:line_ratios}(d) highlights the anti-correlation between \heiia~and \niiia. At high ionization, \heiia~dominates, whereas at lower ionization, \niiia~becomes stronger. This behavior arises naturally from the Bowen fluorescence mechanism, in which \heii~and \niii~share a common excitation pathway.
\end{itemize}

\subsubsection{Emission Line Widths}
\label{sec:linewidth}

\begin{figure}
    \centering
    \includegraphics[width=0.45\linewidth]{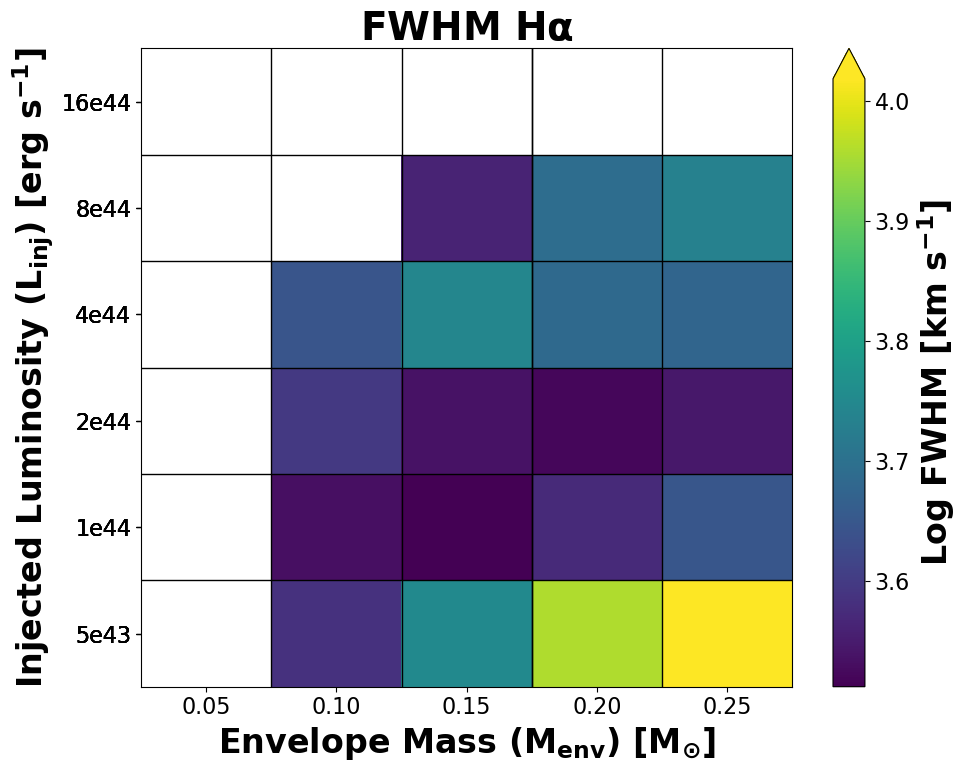}
    \caption{Line widths of \ha. The full width at half maximum (FWHM) of \ha~is shown across the $L_{\rm inj}$--$M_{\rm env}$ parameter space. Our model reproduces the key observational trend: the \ha~line is broadest in H-dominated TDEs and narrowest in spectra exhibiting strong Bowen (\niii) emission.}
    \label{fig:FWHM}
\end{figure}

A defining characteristic of TDE optical emission lines is their broad profiles, with observed full widths at half maximum (FWHM) reaching $\sim 10^4~\rm km~s^{-1}$. Notably, the \ha~line width correlates with spectroscopic class: H-dominated TDEs exhibit systematically broader \ha~lines ($v_{\rm FWHM} \gtrsim 10^4~\rm km~s^{-1}$) than those with prominent Bowen features ($v_{\rm FWHM} \approx \text{few} \times 10^3~\rm km~s^{-1}$) \citep{Charalampopoulos2022}. It has been proposed that this extreme broadening is primarily driven by photon scattering in optically thick gas, rather than bulk rotational motion \citep{Roth17,Parkinson22}. This scenario is directly testable in our outflow model, where the gas has no intrinsic rotation. Any line broadening in our simulations must therefore arise from scattering within the optically thick wind.

The resulting \ha~line widths are shown in Figure~\ref{fig:FWHM}. In the region of parameter space where \ha~is the dominant line (i.e., at the lowest ionization levels), the lines are broadest, reaching $v_{\rm FWHM} \approx 10^4~\rm km~s^{-1}$. In contrast, the narrowest \ha~lines originate in the same region that produces the strongest Bowen \niii~emission, consistent with observations. 

At fixed injected luminosity, the line width generally increases with envelope mass, reflecting the larger electron-scattering optical depth. However, the broadening is determined primarily by the scattering optical depth exterior to the line-forming region. Because the location of this effective photosphere depends sensitively on the radiative transfer, no simple monotonic trend is expected across the full parameter space. 

We note that the \ha~line widths obtained in our simulations likely represent lower limits, as additional broadening mechanisms such as turbulence or gas rotation are not included. Furthermore, envelopes more massive than $0.25\,M_\odot$ may produce even broader lines due to increased photon scattering. While this is a plausible extension, the computational cost of modeling such massive envelopes prevents us from exploring this regime here, leaving this prediction for future work.

\subsection{Physical Mechanisms Producing the Four TDE Spectroscopic Types} \label{sec:spectra_model}
In this section, we synthesize our simulation results to identify the physical conditions and radiative transfer processes that give rise to the four primary TDE spectroscopic types: featureless, He-dominated, Bowen N-dominated, and H-dominated.

Before discussing each class in detail, we first emphasize that the \ha, \hb, and \heiia~emission lines in our simulations are produced predominantly through radiative processes. We briefly summarize the formation mechanisms of the three most prominent features:

\begin{itemize}
    \item \heiia: Under the intense ionizing radiation in our model, helium is primarily in the \heiii~(fully ionized) and \heii~(singly ionized) states. The \heiia~line arises from the $n=4 \rightarrow 3$ transition of \heii. 
    At the temperatures considered here, the line is likely produced by radiative processes, such as recombination and radiative excitation.

    \item \ha~and \hb: The hydrogen Balmer lines are likewise dominated by radiative processes. Photoionization by the central X-ray source is followed by radiative recombination cascades, which populate excited levels. Transitions from $n=3$ and $n=4$ to $n=2$ then produce H$\alpha$ and H$\beta$, respectively. The high ionization rates ensure a steady-state population of excited hydrogen atoms, sustaining strong Balmer emission.

    \item Bowen \niii~lines: The strong \niiia~and \niiib~lines relative to LTE expectations are explained by the Bowen fluorescence mechanism \citep{Bowen1934}. This process is initiated by \helya~photons exciting \oiii. The subsequent decay of \oiii~produces emission that is resonant with transitions in \niii, thereby exciting it. The final radiative decay of \niii~gives rise to the characteristic Bowen lines.
\end{itemize}

The line intensities are set by the upper-level population (i.e., the ionization state), which depend on the radiation field and the gas temperature. In general, a lower radiation field and temperature decrease the ionization rate and increases the recombination rate, enhancing line emission until the parent ion fully recombines. Beyond this point, the corresponding line can no longer be produced, leading to a sharp decline in its intensity.

Motivated by this picture, we analyze the radial ionization structure of hydrogen, helium, and nitrogen throughout the envelope. Because the envelope is optically thick, photons generated in the dense inner regions are largely absorbed or scattered, with their energy reprocessed into the continuum. As a result, the observable emission lines originate primarily from the outer layers where the gas becomes optically thin. Our analysis shows that the ionization fractions of H, He, and N in these surface regions directly regulate the strengths of the emergent lines and therefore determine the spectroscopic classification.

\begin{figure}
    \centering
    \includegraphics[width=0.5\linewidth]{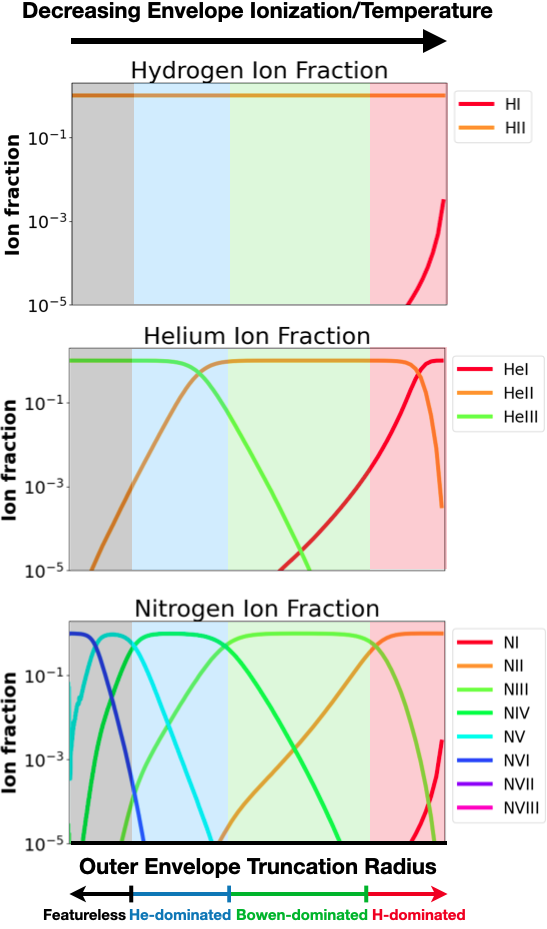}
    \caption{
    Ionization structure of the outer emitting envelope for TDE spectral classes. This schematic shows the ionization fractions of hydrogen (top panel), helium (middle panel), and nitrogen (bottom panel) in the outer layers of the reprocessing envelope, plotted as a function of decreasing ionization level (increasing radius). The curves represent the fractional abundances of the dominant ionization states. The horizontal bar at the base indicates the approximate photospheric radius associated with each spectral class: featureless (grey), He-dominated (blue), Bowen N-dominated (green), and H-dominated (pink).
    } 
    \label{fig:detailed_structure}
\end{figure}

Figure~\ref{fig:detailed_structure} provides a schematic summary of the ionization structure, showing the ionization fractions of H, He, and N as a function of decreasing ionization level (equivalently, increasing radius within the envelope).  The horizontal bar at the base of the plot indicates the location of the outer envelope for each of the four TDE spectral types. A key point is that this plot is not quantitatively comparing which type of TDE has a larger envelope radius (in fact, in our model the photosphere size is fixed) but is meant only to indicate the ionization state at the outer envelope where the lines are produced.

One might deduce from this plot that a more compact photosphere is more likely to produce high-ionization (He-dominated or featureless) spectra. A key point is that the envelope radius does not independently determine the ionization state; rather, one must consider both the envelope mass and the size. For example, when fixing the total envelope mass $M_{\rm env}$, a more compact envelope increases the optical depth in the outer layers, suppressing ionization and favoring lower-ionization spectra. On the other hand, if the envelope central density structure stays unchanged, then shrinking its size reduces optical depth, allowing higher ionization and favoring featureless or \heiia-dominated spectra.

A detailed interpretation of this structure for each spectroscopic class is presented in the following subsections. We emphasize that, although we describe four distinct classes for clarity, the underlying ionization state and resulting spectra vary continuously across the parameter space, leading to natural transitions and mixed cases at the boundaries.

\subsection{Featureless TDEs: Fully ionized regime} \label{sec:featureless}

As indicated by the leftmost grey region in Figure~\ref{fig:detailed_structure}, when the envelope truncates within this ionization zone, the outer gas is dominated by fully ionized hydrogen (\hii) and helium (\heiii). The resulting high ionizing flux and temperature suppress recombination, leading to extremely weak \ha~and \heiia~emission that does not rise above the continuum, and thus produces an observationally featureless spectrum. 

Our exploratory simulations indicate that featureless spectra can also arise through alternative mechanisms. For example, even when emission lines are intrinsically present, sufficiently high wind velocities can broaden them to the point where they become indistinguishable from the continuum. While this suggests that kinematic broadening provides an additional pathway to featureless-like spectra, a detailed investigation of this effect lies beyond the scope of this work, which focuses on photoionization physics, and is left for future study.

\subsubsection{He-dominated TDEs: High ionization state}

As illustrated in Figure~\ref{fig:detailed_structure}, He-dominated TDEs correspond to an envelope that truncates within the blue zone, where a \heiii/\heii~ionization front is present in the outer layers. This front indicates a region where recombination from \heiii~to \heii~is efficient, while the ionizing flux is sufficiently reduced to allow population of excited \heii~levels, leading to strong \heiia~emission.

Because hydrogen has a lower ionization potential, it remains almost fully ionized (\hii) in these outer layers under the same radiation field, resulting in weak Balmer emission and a spectrum dominated by \heiia~over H$\alpha$. Furthermore, nitrogen is primarily in higher ionization states (e.g., \niv, \nv), suppressing the production of observable Bowen (\niii) emission lines.

\subsubsection{Bowen N-dominated TDEs: Moderate ionization state}

A spectrum dominated by Bowen (\niii) emission requires a substantial \niii~fraction in the outer, optically thin layers of the envelope. As shown in Figure~\ref{fig:detailed_structure}, this condition is met when the photosphere lies within the green region, such that the envelope includes the fully ionized (grey) and \heiii-dominated (blue) zones and truncates within the \niii-rich layer. In this configuration, recombination of \heiii~to \heii~at smaller radii provides a source of \helya~photons while also cooling the gas to temperatures favorable for \niii. Within this region, the Bowen fluorescence mechanism becomes efficient, as \helya~photons resonantly excite \oiii, which in turn transfers energy to \niii, producing strong \niiia~and \niiib~emission.

In our simulations, the Bowen \niii~lines can exceed both \ha~and \heiia~in flux. This regime corresponds to moderate ionization and temperatures of order $T \approx 2 \times 10^4$~K in the outer layers. At these temperatures, helium is predominantly in the \heii~state, reducing the recombination rate from \heiii~and thus weakening \heiia~emission. The observable strength of \heiia~also depends on where the line is formed relative to the photosphere: if it originates in optically thin regions, it can escape, whereas if it forms deeper in the envelope, it is suppressed by reprocessing. At the same time, the gas remains sufficiently ionized that hydrogen recombination is inefficient, resulting in weak \ha~emission despite the favorable conditions for \niii.

\subsubsection{Hydrogen-dominated TDEs: Low ionization state}

As the gas temperature and ionization level decrease further, corresponding to the red region in Figure~\ref{fig:detailed_structure}, a \heii/\hei~ionization front develops, enabling more efficient cooling. In this regime, the outer envelope is composed primarily of \hei~and \nii, while hydrogen remains mostly ionized (\hii) with a growing neutral (\hi) fraction. These conditions favor efficient recombination of \hii~to \hi, followed by radiative cascades that produce strong hydrogen Balmer emission.

Because these recombination processes occur in the optically thin outer layers, the resulting H$\alpha$ and H$\beta$ lines emerge prominently and dominate the spectrum. In contrast, \heiia~and Bowen \niii~lines are primarily generated in the hotter inner regions, but are subsequently scattered and absorbed as they propagate through the optically thick outer envelope rich in \hei~and \nii. This suppression of \heiia~at low temperatures is consistent with observational trends, where the coolest TDEs show little to no \heiia~emission \citep[][Fig.~13]{Charalampopoulos2022}.

\subsection{Spectroscopic Evolution with Changing Fallback Rate}
\label{sec:evolution}
Observations indicate that the spectroscopic classification of most TDEs remains relatively stable over time, even as the luminosity declines by an order of magnitude \citep{Charalampopoulos2022}. If the optical emission originates from a reprocessing envelope supplied by fallback debris, and the central luminosity is powered by accretion of that same material, then both the envelope mass $M_{\rm env}$ and the injected luminosity $L_{\rm inj}$ are expected to evolve concurrently.

Although the detailed evolutionary tracks are not known a priori, it is reasonable to assume that both $M_{\rm env}$ and $L_{\rm inj}$ somewhat scale with the fallback rate, under the assumption of approximately constant radiative efficiency and that the envelope mass is regulated by the contemporaneous mass supply. In this scenario, their proportional decline leads to a partial cancellation of effects: the reduction in reprocessing material is offset by the diminishing ionizing flux. As a result, the ionization state varies only weakly with time, and the spectroscopic type remains preserved.

\begin{figure}
    \centering
    \includegraphics[width=0.5\linewidth]{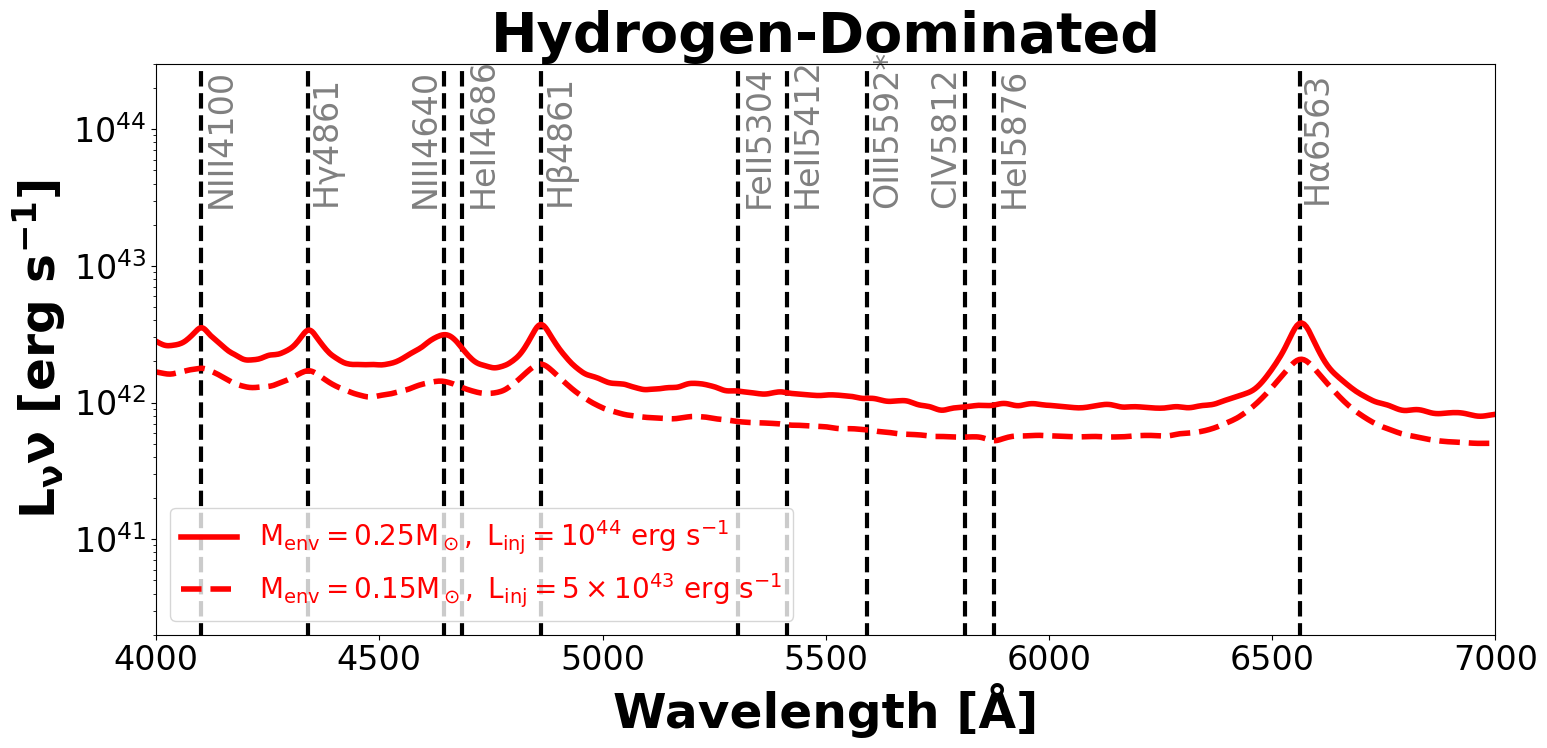}
\includegraphics[width=0.5\linewidth]{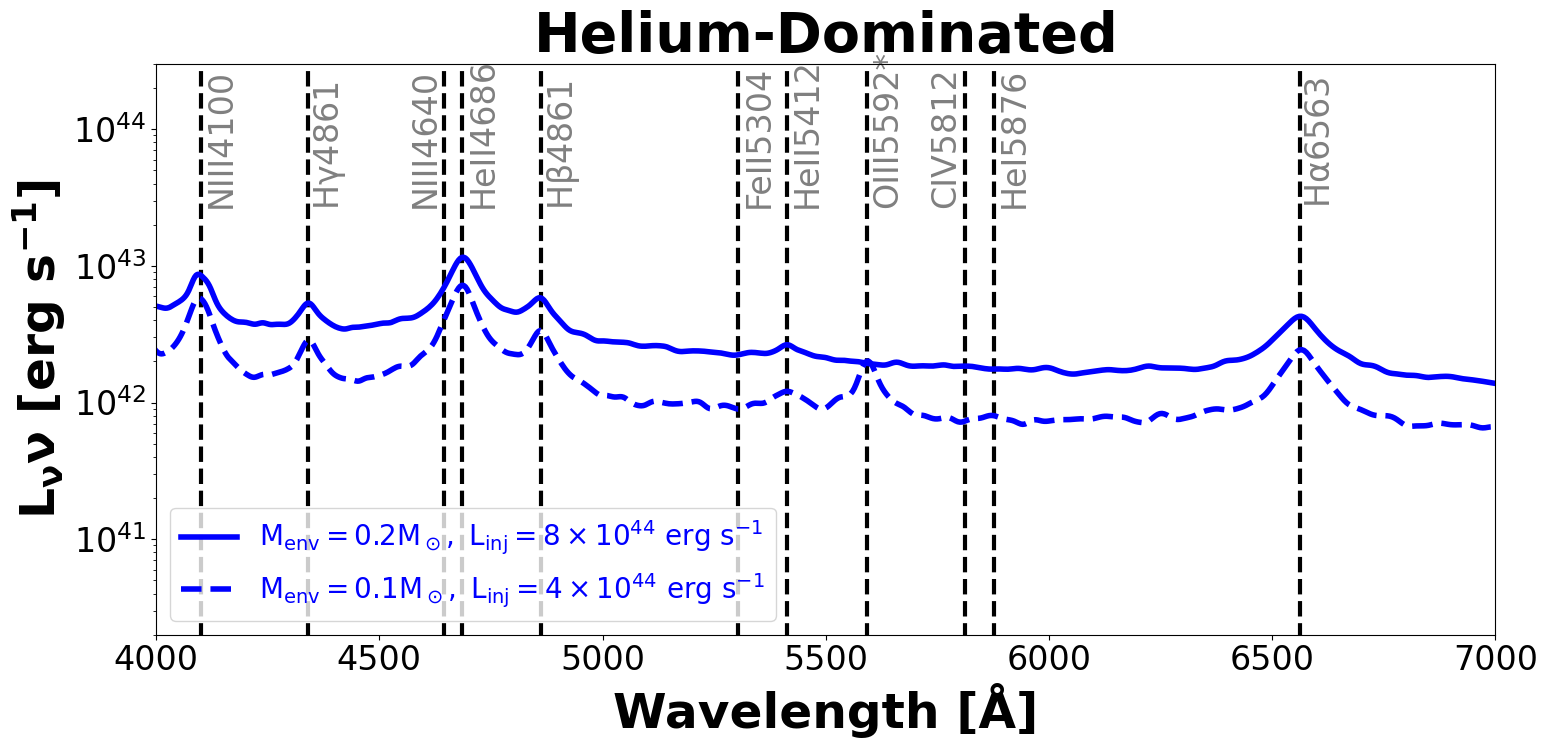}
\includegraphics[width=0.5\linewidth]{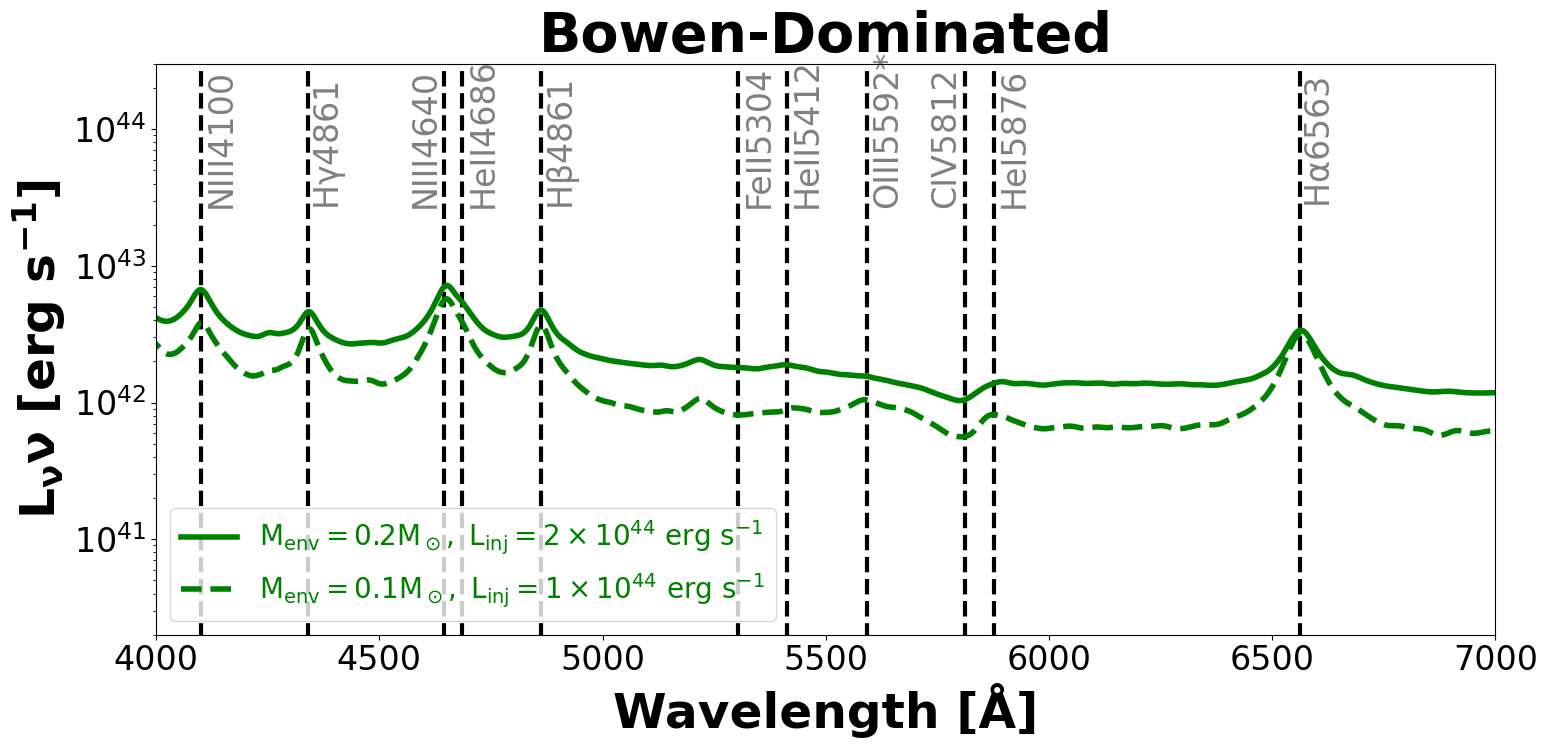}
    \caption{Spectral evolution of TDEs under coupled envelope mass and luminosity decay. The figure shows representative evolutionary tracks across our simulation grid, illustrating that when the injected X-ray luminosity $L_{\rm inj}$ and envelope mass $M_{\rm env}$ decrease proportionally, the effective ionization parameter remains nearly constant, preserving the spectroscopic characteristics over time. The panels display the evolution for H-dominated (top), He-dominated (middle), and Bowen N-dominated (bottom) classes. In each case, the continuum level declines while the relative line strengths and overall spectral shape remain largely unchanged. }
    \label{fig:evol}
\end{figure}

Figure~\ref{fig:evol} illustrates representative evolutionary tracks in the $L_{\rm inj}$--$M_{\rm env}$ parameter space for H-dominated, He-dominated, and Bowen N-dominated TDEs. Along these tracks, the continuum shape and relative line strengths remain largely unchanged, with the primary evolution being an overall decline in luminosity. This behavior naturally explains how a TDE can maintain a stable spectroscopic classification over a significant fraction of its observable lifetime.

We therefore conclude that a near-linear, coupled decline in both luminosity and envelope mass, as expected in fallback-driven evolution, provides a natural explanation for the observed spectral stability. In this simplified picture, other model parameters such as the velocity profile, injected radiation temperature, density slope, and envelope extent are held fixed. In reality, these properties may also evolve as the fallback rate decreases. Although their effects are likely secondary to variations in $M_{\rm env}$ and $L_{\rm inj}$, they may introduce additional spectral diversity at late times.

\subsection{Fitting Observed TDE Spectra with the Reprocessing Model}
\label{sec:fitting}

\begin{figure}
    \centering
    \includegraphics[width=0.45\linewidth]{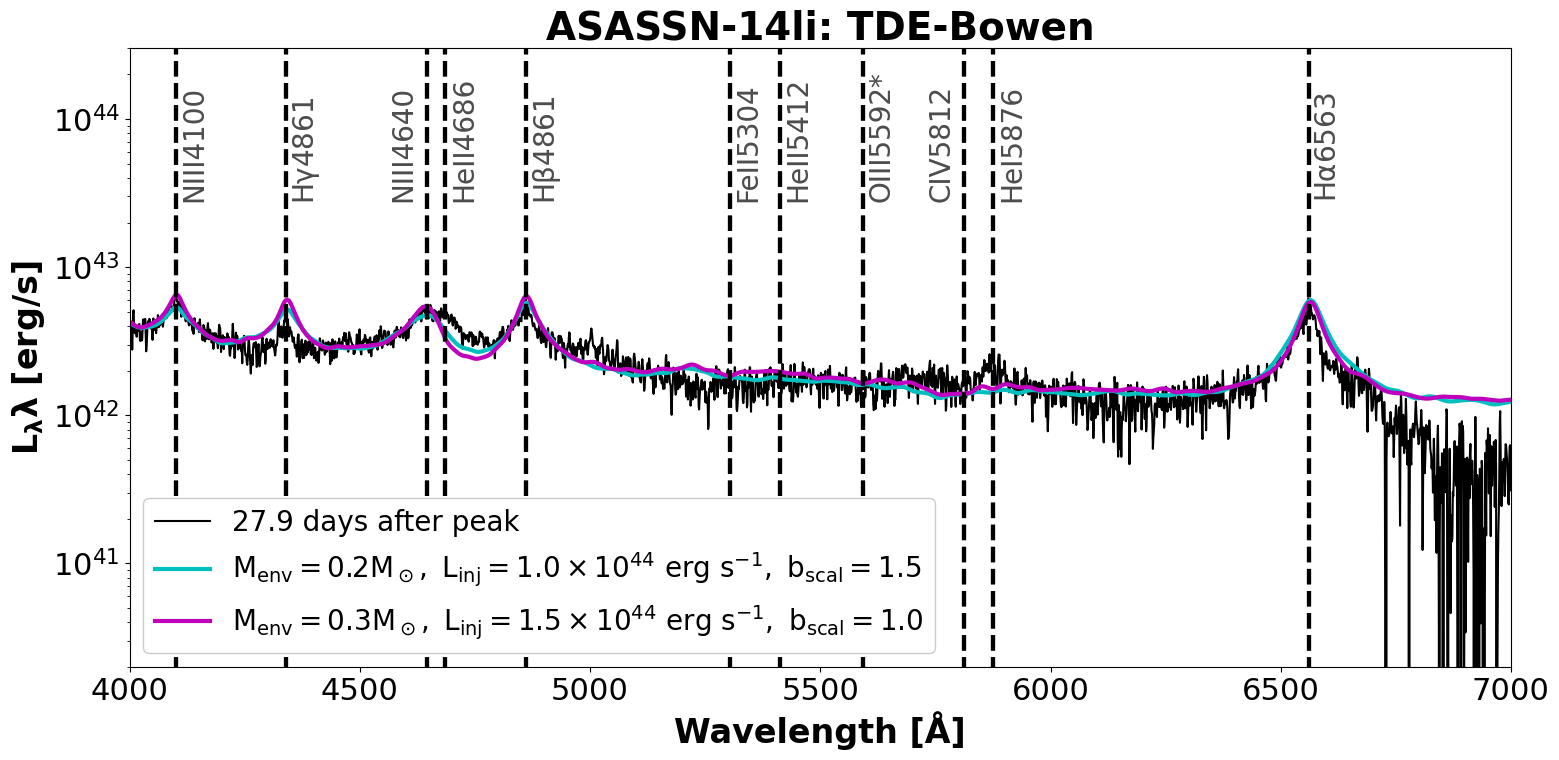}
    \includegraphics[width=0.45\linewidth]{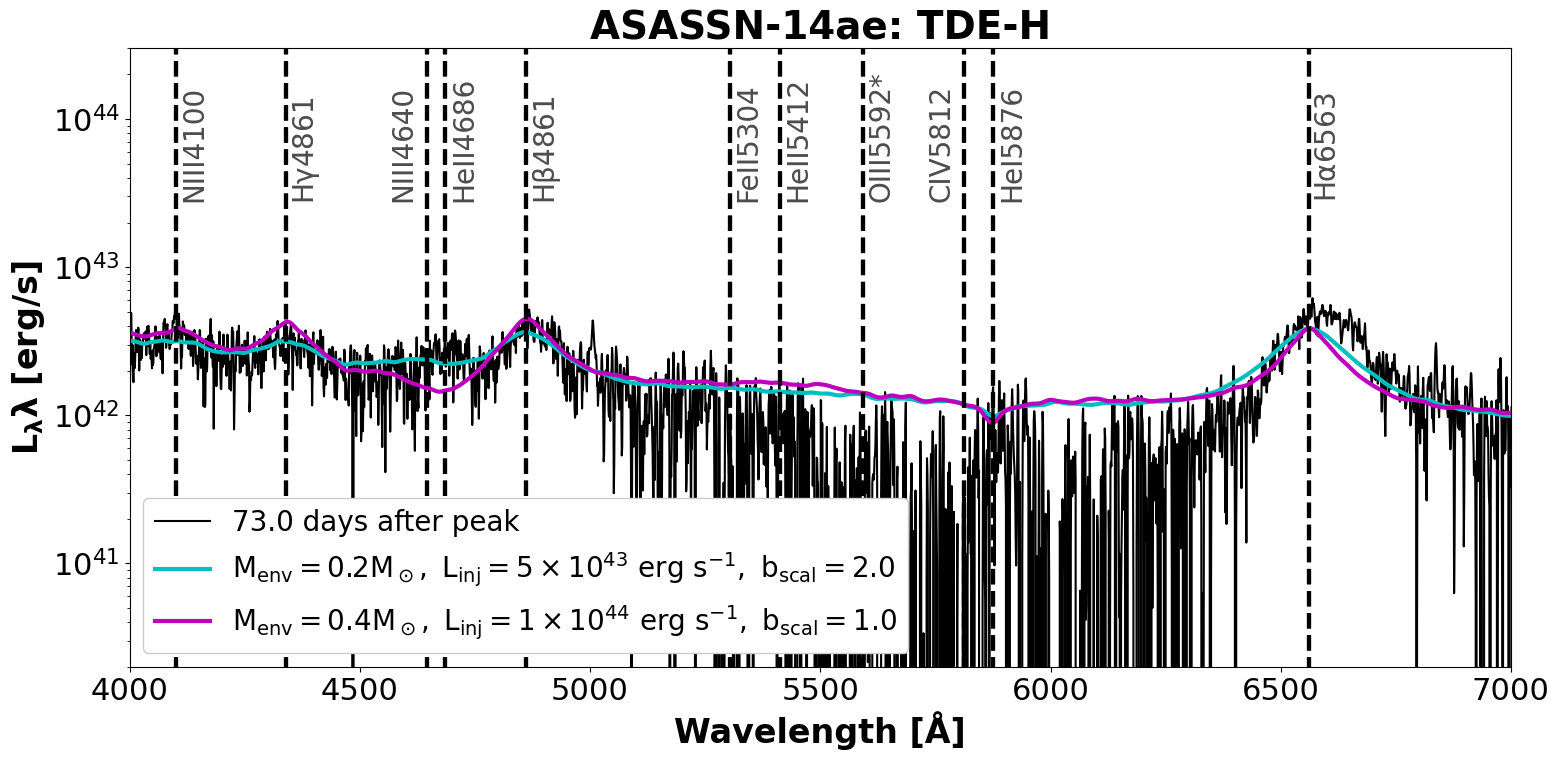}
    \caption{ 
    Fitting observed TDE spectra with the reprocessing model. Observed spectra (black) are compared with simulations from our grid. The magenta curves show the best morphological match, reproducing the spectral type and continuum shape, with the luminosity scaled by a factor $b_{\rm scal}$ for visual comparison. The blue curves correspond to new simulations in which both the envelope mass $M_{\rm env}$ and injected luminosity $L_{\rm inj}$ are scaled by the same factor to match the observed continuum flux.} 
    \label{fig:TDE_obs}
\end{figure}

We apply our model to observed TDE spectra to assess its ability to reproduce the defining features of the main spectroscopic classes. Our comparison uses optical, continuum-subtracted spectra from \citet{Holoien2014,Holoien16_14li,Charalampopoulos2022}. We emphasize that our simulated spectra are drawn from a fiducial envelope model with a fixed parameter grid and have not been fine-tuned for any specific event. Despite this, as shown in Figure~\ref{fig:TDE_obs}, the one-dimensional reprocessing model can reproduce different classes of observed TDEs, capturing both the continuum shape and key spectroscopic features.

In several cases, the continuum level in our fiducial simulations is lower than the observed level. To address this, we introduce a scaling factor, $b_{\text{scal}}$, as shown as the cyan line in Figure~\ref{fig:TDE_obs}. As a proof of concept, this scaling is physically motivated by increasing both the envelope mass ($M_{\text{env}}$) and the injected luminosity ($L_{\text{inj}}$) proportionally to $b_{\text{scal}}$. This joint scaling preserves the ratio $L_{\text{inj}} / M_{\text{env}}$, thereby maintaining a similar ionization level across the envelope while successfully boosting the overall flux level to match the observations (represented by the magenta spectra in Figure~\ref{fig:TDE_obs}).

Our model successfully reproduces both the continuum and the dominant spectroscopic features for several H-dominated and Bowen-dominated TDEs, with good agreement in relative line strengths. It also captures the observed trend in line widths, producing broader lines in TDEs without Bowen features and narrower lines in Bowen-rich systems \citep{Charalampopoulos2022}.

For the two TDEs ASASSN-14ae and ASASSN-14li compared to our models, discrepancies remain near the \hei~and \civ~features around 5800~\AA. These differences are likely due to limitations in the atomic data used in our simulations. Test runs with a more complete atomic dataset show that the \hei~feature transitions from absorption to emission and that the non-physical \oiii~line at $\lambda 5592$ disappears (Appendix~\ref{App:atom}).

\section{Discussion}
\label{sec:discussion}

\subsection{Explaining the Correlation Between TDE Demographics and Spectroscopic Class}
\label{sec:dis:correlation}

Recent systematic studies have identified two observational trends that correlate with TDE spectroscopic class \citep{Velzen2021,Hammerstein2023}. First, there is a luminosity sequence: the mean peak optical/UV luminosity is highest for TDE-featureless events, followed by TDE-He, with TDE-H+He and TDE-H exhibiting progressively lower luminosities. Second, there is a corresponding black hole mass trend: TDE-featureless events are preferentially found in the most massive galaxies (and thus associated with the most massive black holes), followed by TDE-He and TDE-H+He, while TDE-H events are linked to the least massive systems.
We note, however, that these correlations exhibit significant scatter and should be interpreted as broad statistical tendencies rather than strict relationships.

Our reprocessing model naturally accounts for these observed trends. In our framework, the spectroscopic class is set by the ionization state. As shown in Section~\ref{sec:3.1}, the observed UV/optical luminosity also increases with $L_{\rm inj}$ (though not necessarily linearly). A correlation between peak luminosity and spectroscopic class is therefore a direct outcome of the model.

The dependence on black hole mass arises from two coupled effects. First, as the physical size of the envelope likely increases with black hole mass ($R_g \propto M_{\rm BH}$), this distributes the envelope mass  over a larger volume and lowers the optical depth. Second, if the central engine operates near the Eddington limit, the injected luminosity scales as $L_{\rm inj} \propto M_{\rm BH}$. As a result, TDEs around more massive black holes are pushed toward higher ionization states, favoring featureless or He-dominated spectra, in agreement with observations. Conversely, lower-mass black holes yield lower ionization, promoting conditions that produce H-dominated or Bowen N-dominated spectra.

\subsection{Influence of Other Model Parameters}
The primary aim of this work has been to establish the dominant role of the gas ionization state in governing the major spectroscopic classes of TDEs. While this focus provides a unifying framework, other model parameters, such as the gas velocity structure (particularly the terminal velocity), the size and density slope of the envelope, and non-solar elemental abundances, can introduce important secondary effects. A systematic investigation of these factors lies beyond the scope of this foundational study and will be addressed in future work. Here, we highlight several illustrative examples based on our exploratory test runs.

In our fiducial simulation grid (Figure~\ref{fig:TDE_Phase}), we do not obtain spectra that are purely \heiia-dominated without also producing weak \ha~emission. This behavior is consistent with observations, where most TDE-He spectra exhibit detectable, albeit weak, Balmer lines \citep{Charalampopoulos2022}. Nevertheless, we find that near-complete suppression of \ha~while maintaining strong \heiia~can be achieved by adjusting specific parameters. In particular, as demonstrated by \citet{Roth16} and discussed in Appendix~\ref{sec:appendix_He}, a very compact photosphere effectively suppresses hydrogen recombination lines, allowing the model to approach a nearly pure \heii~spectrum.

Second, high terminal velocities can blend nearby emission features, such as \heiia, \niiia, and \hb, into a single broadened component. A substantial fraction of observationally classified TDE-He events (e.g., ASASSN-15oi) exhibit broad and often blueshifted \heiia~profiles, which may arise from this blending effect rather than from a purely He-dominated spectrum.

Finally, a significantly higher terminal velocity than our fiducial value can produce featureless-like spectra. Even when emission lines are intrinsically present, strong scattering in a high-velocity outflow can broaden them to the point that they become indistinguishable from the continuum, leading to an observational classification as TDE-featureless.

\subsection{Linking to the TDE Viewing-Angle-Dependent Unification Model}
\label{sec:discussion:viewing}

Realistic TDE reprocessing envelopes are unlikely to be spherically symmetric. For example, \citet{Dai18,Thomsen2022} proposed that optically thick winds from super-Eddington accretion flows act as reprocessing layers for X-ray photons from the inner disk. In this picture, the density and geometry of the reprocessing material depend strongly on the observer's inclination angle: slower, denser winds are launched at higher inclinations, while faster, more dilute outflows occupy lower-inclination regions. Although multidimensional effects are important \citep[see][for detailed simulations]{Parkinson25}, a simplified one-dimensional line-of-sight approximation captures the essential behavior. In this framework, high-inclination sightlines correspond to models with larger effective envelope masses $M_{\rm env}$, while low-inclination views correspond to smaller effective $M_{\rm env}$. The resulting ionization state and spectral features are therefore intrinsically viewing-angle dependent.

This geometric picture leads to a natural inclination sequence for the spectroscopic classes. Hydrogen-dominated TDEs are expected to arise preferentially at high inclinations, followed by Bowen N-dominated and He-dominated events, while featureless TDEs correspond to low-inclination sightlines that probe hotter, more dilute regions of the flow. However, the envelope mass and ionization state in real systems are also influenced by the mass of the disrupted star and the black hole, not solely by inclination. As a result, inferring the viewing angle directly from the spectroscopic class is not straightforward, since higher fallback rates, lower black hole masses, and larger inclinations all act to increase the envelope density and reduce the ionization state. Multi-wavelength observations, such as radio emission or the detection of blueshifted spectral features in UV or X-rays, may help to break some of these degeneracies.

Some observational evidence supports this unified framework. \citet{Charalampopoulos2022} found that, despite limited sample sizes, TDE-He events are more likely to exhibit X-ray emission than TDE-H. In addition, several featureless TDEs with strong X-ray emission, including AT2022cmc \citep{Andreoni2022} and AT2021ehb \citep{Yao2022}, are thought to be viewed close to face-on. These findings are consistent with the expectation that low-inclination systems exhibit both higher X-ray-to-optical ratios and featureless or He-dominated spectra.  We caution, however, that accretion flow geometries are likely diverse, and a strict one-to-one mapping between X-ray-to-optical ratio and spectroscopic class should not be expected.

\subsection{Contrasting TDE and AGN Broad-Line Physics}
\label{sec:agn}

The spectroscopic features of TDEs differ markedly from those of typical active galactic nuclei (AGNs), reflecting their distinct physical environments. In general, AGN do not exhibit strong \heiia~or Bowen \niii~emission.  The broad \heiia~line is typically weak, with a flux ratio 
$\lesssim 2\%$ of \hb~in composite quasar spectra \citep{VandenBerk2001}, and is often absent in individual objects \citep{Shirazi2012,Kouroumpatzakis2025}. 
However, exceptions exist. For example, recent work has highlighted a class of AGN with strong, variable \heii~lines possibly linked to changing EUV continuum states \citep{MacLeod2019, Lu2025}.

In the classical AGN broad-line region (BLR) model, line emission arises from discrete, gravitationally bound clouds with bulk Keplerian or virialized motions, and line widths are dominated by Doppler broadening \citep{Laor2006,Popovic2006,Kollatschny2013}. However, some lines of evidence suggest a more continuous or outflowing structure, particularly for high-ionization lines like C,\textsc{iv} $\lambda$1549 \citep{Arav1997, Arav1998, Laor2006}.  Nevertheless, even in such wind models, the line broadening remains primarily Doppler due to bulk motion, not electron scattering.

In contrast, in our TDE model, optical emission lines originate from a continuous, optically thick, outflowing wind, where line broadening is dominated by electron scattering within the extended envelope, consistent with previous radiative transfer studies \citep{Roth16,Roth17,Parkinson20,Parkinson22,Parkinson25}. This fundamental difference in the line-broadening mechanism allows TDEs to exhibit significantly broader emission lines than typical AGN.

The contrast in line ratios between AGN and TDEs is primarily driven by differences in ionization state.  In most AGN, the BLR resides at larger radii with insufficient ionization to maintain a substantial \heiii~population, which is required to produce \heiia~through recombination. In contrast, the dense, hot outflows in TDEs sustain extended \heiii~zones, enabling efficient \heii~recombination. These conditions not only produce strong \heiia~emission but also facilitate the Bowen fluorescence mechanism.

Notably, Bowen \niii~lines have recently been detected in transient, high-luminosity flares from a subclass of AGN known as Bowen-fluorescent flares (BFFs) \citep{Trakhtenbrot2019,Frederick2021}. This indicates that the environments of BFFs likely share key characteristics with TDEs, including enhanced EUV or X-ray radiation fields, ionization conditions favorable for the Bowen cascade, and the presence of outflows that enable the necessary resonance processes. However, the emission lines in BFFs are generally narrower than those in TDEs. We interpret this as a difference in the dominant emission region: All known BFFs occur in previously active SMBHs where a classical BLR is already present. In these objects, Bowen fluorescence is activated  within the pre-existing, Doppler-broadened BLR, whereas in TDEs the Bowen lines emerge from a dense, electron-scattering-dominated wind. Consequently, the broader lines in TDEs reflect a higher effective scattering optical depth in the reprocessing material. Future work combining radiative transfer simulations and higher-resolution spectroscopy will be needed to assess this proposal.

\section{Conclusion}
\label{sec:conclusion}
In this work, we have systematically explored the origin of spectroscopic diversity in optical TDEs through a suite of radiative transfer simulations based on a one-dimensional reprocessing envelope model. Our primary findings are as follows:

\begin{itemize}
    \item We demonstrate that the four spectroscopic classes of TDEs (featureless, He-dominated, Bowen N-dominated, and H-dominated) form a continuous sequence governed by a single physical parameter: the gas ionization state. This state is controlled by the balance between the injected X-ray luminosity $L_{\rm inj}$ and the mass of the surrounding outflowing envelope $M_{\rm env}$. A smooth progression across the $L_{\rm inj}$--$M_{\rm env}$ parameter space naturally reproduces the full observed sequence.

    \item Our simulations show that optical emission lines in TDEs arise from radiative processes in an optically thick, super-Eddington wind. A key result is that the extreme line widths (FWHM $\approx 10^4$ km~s$^{-1}$) arise from electron scattering within this wind, in contrast to the kinematic broadening that dominates AGN broad-line regions. The model reproduces the observed trend that H$\alpha$ lines are broader in H-dominated TDEs and narrower in Bowen N-dominated events, linking line width directly to the wind optical depth.

    \item The model provides a natural explanation for observed correlations between spectroscopic class and host properties. The scaling of envelope density and ionization with black hole mass drives TDEs around more massive black holes toward higher ionization states, favoring featureless or He-dominated spectra and their associated higher peak luminosities. In addition, the model explains the observed spectral stability within individual TDEs: if $L_{\rm inj}$ and $M_{\rm env}$ decline proportionally with the debris fallback rate, the ionization parameter remains approximately constant, preserving the spectroscopic class even as the total luminosity fades.

\end{itemize}

As this work was being prepared, another work by Asperen \& Kasen (2026) also conducted Sedona calculations to show that high luminosity and compact radii in transients like LFBOTs and TDEs produce featureless spectra by raising the gas temperature and ionization state. Although their model setup and radiative transfer physics treatment are different from our study, the results are broadly consistent. This illustrates that the physical origin of spectroscopic diversity in different types of transients could be universal, governed by the same fundamental principles of radiation--gas interaction irrespective of the specific progenitor or transient class. The qualitative trend that higher ionization favors helium-dominated or featureless spectra while lower ionization favors hydrogen-dominated spectra is robust across different methods, including our non-LTE treatment and the LTE approach of Asperen \& Kasen (2026), as well as previous observational and theoretical studies.

In summary, this study shows that the diverse spectroscopic classes of optical TDEs can be largely understood as a consequence of variations in the gas ionization state within an optically thick, outflowing envelope. Our results provide a unified physical interpretation of TDE spectra, explain key observational trends, and establish a practical framework for spectral modeling. Future work exploring additional factors, including non-solar abundances, detailed velocity and density structures, and multidimensional geometry, will further refine this picture and deepen our understanding of these extreme transient events.

\begin{acknowledgments}
We thank N. Jiang, N. Roth, B. Trakhtenbrot, Y. Yao for useful discussions. L.T. and L.D. acknowledge support from the Hong Kong Research Grants Council (RGC GRF 17305124), the National Natural Science Foundation of China (NSFC/RGC JRS N\_HKU782/23),  and the National Key R\&D Program of China (2025YFF0511100). E.R-R. is supported in part by the Heising-Simons Foundation, the Vera Rubin Presidential Chair at UCSC, and the National Science Foundation (AST-2307710, AST-2206243, AST-1911206). P.C. acknowledges financial support from the Secretary of Universities and Research (Government of Catalonia) and by the Horizon 2020 Research and Innovation Programme of the European Union under the Marie Sk\l{}odowska-Curie and the Beatriu de Pin\'os 2024 BP 00125 programme, and from the Centro Superior de Investigaciones Cient\'ificas (CSIC) under the Spanish program Unidad de Excelencia María de Maeztu CEX2020-001058-M, financed by MCIN/AEI/10.13039/501100011033, and by the MaX-CSIC Excellence Award MaX4-SOMMA-ICE. G. L. was supported by a research grant (60862) from VILLUM FONDEN.
\end{acknowledgments}

\bibliographystyle{yahapj}
\bibliography{references}

@ARTICLE{Bush2025,
       author = {{Bush}, Rewa Clark and {Wu}, Samantha C. and {Everson}, Rosa Wallace and {Yarza}, Ricardo and {Murguia-Berthier}, Ariadna and {Ramirez-Ruiz}, Enrico},
        title = "{Black Hole Survival Guide: Searching for Stars in the Galactic Center that Endure Partial Tidal Disruption}",
      journal = {\apjl},
         year = 2025,
        month = sep,
       volume = {990},
       number = {1},
          eid = {L7},
        pages = {L7},
          doi = {10.3847/2041-8213/adefde},
archivePrefix = {arXiv},
       eprint = {2504.14705},
 primaryClass = {astro-ph.SR},
       adsurl = {https://ui.adsabs.harvard.edu/abs/2025ApJ...990L...7B}
}

@ARTICLE{Arav1997,
       author = {{Arav}, Nahum and {Barlow}, Tom A. and {Laor}, Ari and {Blandford}, Roger D.},
        title = "{Keck high-resolution spectroscopy of MRK 335: constraints on the number of emitting clouds in the broad-line region}",
      journal = {\mnras},
         year = 1997,
        month = jul,
       volume = {288},
       number = {4},
        pages = {1015-1021},
          doi = {10.1093/mnras/288.4.1015},
       adsurl = {https://ui.adsabs.harvard.edu/abs/1997MNRAS.288.1015A}
}

@ARTICLE{Arav1998,
       author = {{Arav}, Nahum and {Barlow}, Tom A. and {Laor}, Ari and {Sargent}, Wallace L.~W. and {Blandford}, Roger D.},
        title = "{Are AGN broad emission lines formed by discrete clouds? Analysis of Keck high-resolution spectroscopy of NGC 4151}",
      journal = {\mnras},
         year = 1998,
        month = jul,
       volume = {297},
       number = {4},
        pages = {990-998},
          doi = {10.1046/j.1365-8711.1998.297004990.x},
archivePrefix = {arXiv},
       eprint = {astro-ph/9801012},
 primaryClass = {astro-ph},
       adsurl = {https://ui.adsabs.harvard.edu/abs/1998MNRAS.297..990A}
}

@ARTICLE{MacLeod2019,
       author = {{MacLeod}, Chelsea L. and {Green}, Paul J. and {Anderson}, Scott F. and {Bruce}, Alastair and {Eracleous}, Michael and {Graham}, Matthew and {Homan}, David and {Lawrence}, Andy and {LeBleu}, Amy and {Ross}, Nicholas P. and {Ruan}, John J. and {Runnoe}, Jessie and {Stern}, Daniel and {Burgett}, William and {Chambers}, Kenneth C. and {Kaiser}, Nick and {Magnier}, Eugene and {Metcalfe}, Nigel},
        title = "{Changing-look Quasar Candidates: First Results from Follow-up Spectroscopy of Highly Optically Variable Quasars}",
      journal = {\apj},
         year = 2019,
        month = mar,
       volume = {874},
       number = {1},
          eid = {8},
        pages = {8},
          doi = {10.3847/1538-4357/ab05e2},
archivePrefix = {arXiv},
       eprint = {1810.00087},
 primaryClass = {astro-ph.GA},
       adsurl = {https://ui.adsabs.harvard.edu/abs/2019ApJ...874....8M}
}

@ARTICLE{Lu2025,
       author = {{Lu}, Wen-Tao and {Wang}, Jun-Xian},
        title = "{A Sample of He II {\ensuremath{\lambda}}4686 ``Changing-look'' Quasars}",
      journal = {\apj},
         year = 2025,
        month = dec,
       volume = {994},
       number = {2},
          eid = {154},
        pages = {154},
          doi = {10.3847/1538-4357/ae11a2},
archivePrefix = {arXiv},
       eprint = {2510.06753},
 primaryClass = {astro-ph.GA},
       adsurl = {https://ui.adsabs.harvard.edu/abs/2025ApJ...994..154L}
}

@ARTICLE{VandenBerk2001,
       author = {{Vanden Berk}, Daniel E. and {Richards}, Gordon T. and {Bauer}, Amanda and {Strauss}, Michael A. and {Schneider}, Donald P. and {Heckman}, Timothy M. and {York}, Donald G. and {Hall}, Patrick B. and {Fan}, Xiaohui and {Knapp}, G.~R. and {Anderson}, Scott F. and {Annis}, James and {Bahcall}, Neta A. and {Bernardi}, Mariangela and {Briggs}, John W. and {Brinkmann}, J. and {Brunner}, Robert and {Burles}, Scott and {Carey}, Larry and {Castander}, Francisco J. and {Connolly}, A.~J. and {Crocker}, J.~H. and {Csabai}, Istv{\'a}n and {Doi}, Mamoru and {Finkbeiner}, Douglas and {Friedman}, Scott and {Frieman}, Joshua A. and {Fukugita}, Masataka and {Gunn}, James E. and {Hennessy}, G.~S. and {Ivezi{\'c}}, {\v{Z}}eljko and {Kent}, Stephen and {Kunszt}, Peter Z. and {Lamb}, D.~Q. and {Leger}, R. French and {Long}, Daniel C. and {Loveday}, Jon and {Lupton}, Robert H. and {Meiksin}, Avery and {Merelli}, Aronne and {Munn}, Jeffrey A. and {Newberg}, Heidi Jo and {Newcomb}, Matt and {Nichol}, R.~C. and {Owen}, Russell and {Pier}, Jeffrey R. and {Pope}, Adrian and {Rockosi}, Constance M. and {Schlegel}, David J. and {Siegmund}, Walter A. and {Smee}, Stephen and {Snir}, Yehuda and {Stoughton}, Chris and {Stubbs}, Christopher and {SubbaRao}, Mark and {Szalay}, Alexander S. and {Szokoly}, Gyula P. and {Tremonti}, Christy and {Uomoto}, Alan and {Waddell}, Patrick and {Yanny}, Brian and {Zheng}, Wei},
        title = "{Composite Quasar Spectra from the Sloan Digital Sky Survey}",
      journal = {\aj},
         year = 2001,
        month = aug,
       volume = {122},
       number = {2},
        pages = {549-564},
          doi = {10.1086/321167},
archivePrefix = {arXiv},
       eprint = {astro-ph/0105231},
 primaryClass = {astro-ph},
       adsurl = {https://ui.adsabs.harvard.edu/abs/2001AJ....122..549V}
}

@ARTICLE{Mockler2024,
       author = {{Mockler}, Brenna and {Gallegos-Garcia}, Monica and {G{\"o}tberg}, Ylva and {Miller}, Jon M. and {Ramirez-Ruiz}, Enrico},
        title = "{Tidal Disruption Events from Stripped Stars}",
      journal = {\apjl},
         year = 2024,
        month = sep,
       volume = {973},
       number = {1},
          eid = {L9},
        pages = {L9},
          doi = {10.3847/2041-8213/ad6c34},
archivePrefix = {arXiv},
       eprint = {2406.04455},
 primaryClass = {astro-ph.HE},
       adsurl = {https://ui.adsabs.harvard.edu/abs/2024ApJ...973L...9M}
}

@ARTICLE{Mockler2022,
       author = {{Mockler}, Brenna and {Twum}, Angela A. and {Auchettl}, Katie and {Dodd}, Sierra and {French}, K.~D. and {Law-Smith}, Jamie A.~P. and {Ramirez-Ruiz}, Enrico},
        title = "{Evidence for the Preferential Disruption of Moderately Massive Stars by Supermassive Black Holes}",
      journal = {\apj},
         year = 2022,
        month = jan,
       volume = {924},
       number = {2},
          eid = {70},
        pages = {70},
          doi = {10.3847/1538-4357/ac35d5},
archivePrefix = {arXiv},
       eprint = {2110.03013},
 primaryClass = {astro-ph.HE},
       adsurl = {https://ui.adsabs.harvard.edu/abs/2022ApJ...924...70M}
}

@ARTICLE{Auchettl2017,
       author = {{Auchettl}, Katie and {Guillochon}, James and {Ramirez-Ruiz}, Enrico},
        title = "{New Physical Insights about Tidal Disruption Events from a Comprehensive Observational Inventory at X-Ray Wavelengths}",
      journal = {\apj},
         year = 2017,
        month = apr,
       volume = {838},
       number = {2},
          eid = {149},
        pages = {149},
          doi = {10.3847/1538-4357/aa633b},
archivePrefix = {arXiv},
       eprint = {1611.02291},
 primaryClass = {astro-ph.HE},
       adsurl = {https://ui.adsabs.harvard.edu/abs/2017ApJ...838..149A}
}

@ARTICLE{Ramirez-Ruiz2009,
       author = {{Ramirez-Ruiz}, Enrico and {Rosswog}, Stephan},
        title = "{The Star Ingesting Luminosity of Intermediate-Mass Black Holes in Globular Clusters}",
      journal = {\apjl},
         year = 2009,
        month = jun,
       volume = {697},
       number = {2},
        pages = {L77-L80},
          doi = {10.1088/0004-637X/697/2/L77},
archivePrefix = {arXiv},
       eprint = {0808.3847},
 primaryClass = {astro-ph},
       adsurl = {https://ui.adsabs.harvard.edu/abs/2009ApJ...697L..77R}
}

@ARTICLE{Giannios2011,
       author = {{Giannios}, Dimitrios and {Metzger}, Brian D.},
        title = "{Radio transients from stellar tidal disruption by massive black holes}",
      journal = {\mnras},
         year = 2011,
        month = sep,
       volume = {416},
       number = {3},
        pages = {2102-2107},
          doi = {10.1111/j.1365-2966.2011.19188.x},
archivePrefix = {arXiv},
       eprint = {1102.1429},
 primaryClass = {astro-ph.HE},
       adsurl = {https://ui.adsabs.harvard.edu/abs/2011MNRAS.416.2102G}
}

@ARTICLE{DeColle2012,
       author = {{De Colle}, Fabio and {Guillochon}, James and {Naiman}, Jill and {Ramirez-Ruiz}, Enrico},
        title = "{The Dynamics, Appearance, and Demographics of Relativistic Jets Triggered by Tidal Disruption of Stars in Quiescent Supermassive Black Holes}",
      journal = {\apj},
         year = 2012,
        month = dec,
       volume = {760},
       number = {2},
          eid = {103},
        pages = {103},
          doi = {10.1088/0004-637X/760/2/103},
archivePrefix = {arXiv},
       eprint = {1205.1507},
 primaryClass = {astro-ph.HE},
       adsurl = {https://ui.adsabs.harvard.edu/abs/2012ApJ...760..103D}
}

@ARTICLE{Guillochon2014,
       author = {{Guillochon}, James and {Manukian}, Haik and {Ramirez-Ruiz}, Enrico},
        title = "{PS1-10jh: The Disruption of a Main-sequence Star of Near-solar Composition}",
      journal = {\apj},
         year = 2014,
        month = mar,
       volume = {783},
       number = {1},
          eid = {23},
        pages = {23},
          doi = {10.1088/0004-637X/783/1/23},
archivePrefix = {arXiv},
       eprint = {1304.6397},
 primaryClass = {astro-ph.HE},
       adsurl = {https://ui.adsabs.harvard.edu/abs/2014ApJ...783...23G}
}

@ARTICLE{Miller2023,
       author = {{Miller}, Jon M. and {Mockler}, Brenna and {Ramirez-Ruiz}, Enrico and {Draghis}, Paul A. and {Drake}, Jeremy J. and {Raymond}, John and {Reynolds}, Mark T. and {Xiang}, Xin and {Yun}, Sol Bin and {Zoghbi}, Abderahmen},
        title = "{Evidence of a Massive Stellar Disruption in the X-Ray Spectrum of ASASSN-14li}",
      journal = {\apjl},
         year = 2023,
        month = aug,
       volume = {953},
       number = {2},
          eid = {L23},
        pages = {L23},
          doi = {10.3847/2041-8213/ace03c},
archivePrefix = {arXiv},
       eprint = {2308.10964},
 primaryClass = {astro-ph.HE},
       adsurl = {https://ui.adsabs.harvard.edu/abs/2023ApJ...953L..23M}
}

@ARTICLE{Kochanek2016,
       author = {{Kochanek}, C.~S.},
        title = "{Abundance anomalies in tidal disruption events}",
      journal = {\mnras},
         year = 2016,
        month = may,
       volume = {458},
       number = {1},
        pages = {127-134},
          doi = {10.1093/mnras/stw267},
archivePrefix = {arXiv},
       eprint = {1512.03065},
 primaryClass = {astro-ph.HE},
       adsurl = {https://ui.adsabs.harvard.edu/abs/2016MNRAS.458..127K}
}

@ARTICLE{Nicholl2022,
       author = {{Nicholl}, Matt and {Lanning}, Daniel and {Ramsden}, Paige and {Mockler}, Brenna and {Lawrence}, Andy and {Short}, Phil and {Ridley}, Evan J.},
        title = "{Systematic light-curve modelling of TDEs: statistical differences between the spectroscopic classes}",
      journal = {\mnras},
         year = 2022,
        month = oct,
       volume = {515},
       number = {4},
        pages = {5604-5616},
          doi = {10.1093/mnras/stac2206},
archivePrefix = {arXiv},
       eprint = {2201.02649},
 primaryClass = {astro-ph.HE},
       adsurl = {https://ui.adsabs.harvard.edu/abs/2022MNRAS.515.5604N}
}

@ARTICLE{Hu2026,
       author = {{Hu}, Fangyi (Fitz) and {Mandel}, Ilya and {Nealon}, Rebecca and {Price}, Daniel J.},
        title = "{Converged Simulations of the Nozzle Shock in Tidal Disruption Events}",
      journal = {\apjl},
         year = 2026,
        month = jan,
       volume = {996},
       number = {2},
          eid = {L21},
        pages = {L21},
          doi = {10.3847/2041-8213/ae27cc},
archivePrefix = {arXiv},
       eprint = {2510.04790},
 primaryClass = {astro-ph.HE},
       adsurl = {https://ui.adsabs.harvard.edu/abs/2026ApJ...996L..21H}
}

@ARTICLE{Xiaoshan2024,
       author = {{Huang}, Xiaoshan and {Davis}, Shane W. and {Jiang}, Yan-fei},
        title = "{Pre-peak Emission in Tidal Disruption Events}",
      journal = {\apj},
         year = 2024,
        month = oct,
       volume = {974},
       number = {2},
          eid = {165},
        pages = {165},
          doi = {10.3847/1538-4357/ad6c39},
archivePrefix = {arXiv},
       eprint = {2404.18446},
 primaryClass = {astro-ph.HE},
       adsurl = {https://ui.adsabs.harvard.edu/abs/2024ApJ...974..165H}
}

@ARTICLE{Parkinson25,
       author = {{Parkinson}, Edward J. and {Knigge}, Christian and {Dai}, Lixin and {Thomsen}, Lars Lund and {Matthews}, James H. and {Long}, Knox S.},
        title = "{A multidimensional view of a unified model for TDEs}",
      journal = {\mnras},
         year = 2025,
        month = jul,
       volume = {540},
       number = {4},
        pages = {3069-3085},
          doi = {10.1093/mnras/staf880},
archivePrefix = {arXiv},
       eprint = {2408.16371},
 primaryClass = {astro-ph.HE},
       adsurl = {https://ui.adsabs.harvard.edu/abs/2025MNRAS.540.3069P}
}

@ARTICLE{Alexander20review,
       author = {{Alexander}, Kate D. and {van Velzen}, Sjoert and {Horesh}, Assaf and {Zauderer}, B. Ashley},
        title = "{Radio Properties of Tidal Disruption Events}",
      journal = {\ssr},
         year = 2020,
        month = jun,
       volume = {216},
       number = {5},
          eid = {81},
        pages = {81},
          doi = {10.1007/s11214-020-00702-w},
archivePrefix = {arXiv},
       eprint = {2006.01159},
 primaryClass = {astro-ph.HE},
       adsurl = {https://ui.adsabs.harvard.edu/abs/2020SSRv..216...81A}
}

@ARTICLE{Steinberg2024,
       author = {{Steinberg}, Elad and {Stone}, Nicholas C.},
        title = "{Stream-disk shocks as the origins of peak light in tidal disruption events}",
      journal = {\nat},
         year = 2024,
        month = jan,
       volume = {625},
       number = {7995},
        pages = {463-467},
          doi = {10.1038/s41586-023-06875-y},
archivePrefix = {arXiv},
       eprint = {2206.10641},
 primaryClass = {astro-ph.HE},
       adsurl = {https://ui.adsabs.harvard.edu/abs/2024Natur.625..463S}
}

@ARTICLE{Leloudas2019,
       author = {{Leloudas}, Giorgos and {Dai}, Lixin and {Arcavi}, Iair and {Vreeswijk}, Paul M. and {Mockler}, Brenna and {Roy}, Rupak and {Malesani}, Daniele B. and {Schulze}, Steve and {Wevers}, Thomas and {Fraser}, Morgan and {Ramirez-Ruiz}, Enrico and {Auchettl}, Katie and {Burke}, Jamison and {Cannizzaro}, Giacomo and {Charalampopoulos}, Panos and {Chen}, Ting-Wan and {Cikota}, Aleksandar and {Della Valle}, Massimo and {Galbany}, Lluis and {Gromadzki}, Mariusz and {Heintz}, Kasper E. and {Hiramatsu}, Daichi and {Jonker}, Peter G. and {Kostrzewa-Rutkowska}, Zuzanna and {Maguire}, Kate and {Mandel}, Ilya and {Nicholl}, Matt and {Onori}, Francesca and {Roth}, Nathaniel and {Smartt}, Stephen J. and {Wyrzykowski}, Lukasz and {Young}, Dave R.},
        title = "{The Spectral Evolution of AT 2018dyb and the Presence of Metal Lines in Tidal Disruption Events}",
      journal = {\apj},
         year = 2019,
        month = dec,
       volume = {887},
       number = {2},
          eid = {218},
        pages = {218},
          doi = {10.3847/1538-4357/ab5792},
archivePrefix = {arXiv},
       eprint = {1903.03120},
 primaryClass = {astro-ph.HE},
       adsurl = {https://ui.adsabs.harvard.edu/abs/2019ApJ...887..218L}
}

@ARTICLE{Hammerstein2023,
       author = {{Hammerstein}, Erica and {van Velzen}, Sjoert and {Gezari}, Suvi and {Cenko}, S. Bradley and {Yao}, Yuhan and {Ward}, Charlotte and {Frederick}, Sara and {Villanueva}, Natalia and {Somalwar}, Jean J. and {Graham}, Matthew J. and {Kulkarni}, Shrinivas R. and {Stern}, Daniel and {Andreoni}, Igor and {Bellm}, Eric C. and {Dekany}, Richard and {Dhawan}, Suhail and {Drake}, Andrew J. and {Fremling}, Christoffer and {Gatkine}, Pradip and {Groom}, Steven L. and {Ho}, Anna Y.~Q. and {Kasliwal}, Mansi M. and {Karambelkar}, Viraj and {Kool}, Erik C. and {Masci}, Frank J. and {Medford}, Michael S. and {Perley}, Daniel A. and {Purdum}, Josiah and {van Roestel}, Jan and {Sharma}, Yashvi and {Sollerman}, Jesper and {Taggart}, Kirsty and {Yan}, Lin},
        title = "{The Final Season Reimagined: 30 Tidal Disruption Events from the ZTF-I Survey}",
      journal = {\apj},
         year = 2023,
        month = jan,
       volume = {942},
       number = {1},
          eid = {9},
        pages = {9},
          doi = {10.3847/1538-4357/aca283},
archivePrefix = {arXiv},
       eprint = {2203.01461},
 primaryClass = {astro-ph.HE},
       adsurl = {https://ui.adsabs.harvard.edu/abs/2023ApJ...942....9H}
}

@ARTICLE{Thomsen2022,
       author = {{Thomsen}, Lars L. and {Kwan}, Tom M. and {Dai}, Lixin and {Wu}, Samantha C. and {Roth}, Nathaniel and {Ramirez-Ruiz}, Enrico},
        title = "{Dynamical Unification of Tidal Disruption Events}",
      journal = {\apjl},
         year = 2022,
        month = oct,
       volume = {937},
       number = {2},
          eid = {L28},
        pages = {L28},
          doi = {10.3847/2041-8213/ac911f},
archivePrefix = {arXiv},
       eprint = {2206.02804},
 primaryClass = {astro-ph.HE},
       adsurl = {https://ui.adsabs.harvard.edu/abs/2022ApJ...937L..28T}
}

@ARTICLE{Yao2022,
       author = {{Yao}, Yuhan and {Lu}, Wenbin and {Guolo}, Muryel and {Pasham}, Dheeraj R. and {Gezari}, Suvi and {Gilfanov}, Marat and {Gendreau}, Keith C. and {Harrison}, Fiona and {Cenko}, S. Bradley and {Kulkarni}, S.~R. and {Miller}, Jon M. and {Walton}, Dominic J. and {Garc{\'\i}a}, Javier A. and {van Velzen}, Sjoert and {Alexander}, Kate D. and {Miller-Jones}, James C.~A. and {Nicholl}, Matt and {Hammerstein}, Erica and {Medvedev}, Pavel and {Stern}, Daniel and {Ravi}, Vikram and {Sunyaev}, R. and {Bloom}, Joshua S. and {Graham}, Matthew J. and {Kool}, Erik C. and {Mahabal}, Ashish A. and {Masci}, Frank J. and {Purdum}, Josiah and {Rusholme}, Ben and {Sharma}, Yashvi and {Smith}, Roger and {Sollerman}, Jesper},
        title = "{The Tidal Disruption Event AT2021ehb: Evidence of Relativistic Disk Reflection, and Rapid Evolution of the Disk-Corona System}",
      journal = {\apj},
         year = 2022,
        month = sep,
       volume = {937},
       number = {1},
          eid = {8},
        pages = {8},
          doi = {10.3847/1538-4357/ac898a},
archivePrefix = {arXiv},
       eprint = {2206.12713},
 primaryClass = {astro-ph.HE},
       adsurl = {https://ui.adsabs.harvard.edu/abs/2022ApJ...937....8Y}
}

@ARTICLE{Bonnerot21,
       author = {{Bonnerot}, Cl{\'e}ment and {Lu}, Wenbin and {Hopkins}, Philip F.},
        title = "{First light from tidal disruption events}",
      journal = {\mnras},
         year = 2021,
        month = jul,
       volume = {504},
       number = {4},
        pages = {4885-4905},
          doi = {10.1093/mnras/stab398},
archivePrefix = {arXiv},
       eprint = {2012.12271},
 primaryClass = {astro-ph.HE},
       adsurl = {https://ui.adsabs.harvard.edu/abs/2021MNRAS.504.4885B}
}

@ARTICLE{Charalampopoulos2022,
       author = {{Charalampopoulos}, P. and {Leloudas}, G. and {Malesani}, D.~B. and {Wevers}, T. and {Arcavi}, I. and {Nicholl}, M. and {Pursiainen}, M. and {Lawrence}, A. and {Anderson}, J.~P. and {Benetti}, S. and {Cannizzaro}, G. and {Chen}, T. -W. and {Galbany}, L. and {Gromadzki}, M. and {Guti{\'e}rrez}, C.~P. and {Inserra}, C. and {Jonker}, P.~G. and {M{\"u}ller-Bravo}, T.~E. and {Onori}, F. and {Short}, P. and {Sollerman}, J. and {Young}, D.~R.},
        title = "{A detailed spectroscopic study of tidal disruption events}",
      journal = {\aap},
         year = 2022,
        month = mar,
       volume = {659},
          eid = {A34},
        pages = {A34},
          doi = {10.1051/0004-6361/202142122},
archivePrefix = {arXiv},
       eprint = {2109.00016},
 primaryClass = {astro-ph.HE},
       adsurl = {https://ui.adsabs.harvard.edu/abs/2022A&A...659A..34C}
}

@ARTICLE{Coughlin2014,
       author = {{Coughlin}, Eric R. and {Begelman}, Mitchell C.},
        title = "{Hyperaccretion during Tidal Disruption Events: Weakly Bound Debris Envelopes and Jets}",
      journal = {\apj},
         year = 2014,
        month = feb,
       volume = {781},
       number = {2},
          eid = {82},
        pages = {82},
          doi = {10.1088/0004-637X/781/2/82},
archivePrefix = {arXiv},
       eprint = {1312.5314},
 primaryClass = {astro-ph.HE},
       adsurl = {https://ui.adsabs.harvard.edu/abs/2014ApJ...781...82C}
}

@ARTICLE{Dai18,
       author = {{Dai}, Lixin and {McKinney}, Jonathan C. and {Roth}, Nathaniel and
         {Ramirez-Ruiz}, Enrico and {Miller}, M. Coleman},
        title = "{A Unified Model for Tidal Disruption Events}",
      journal = {\apj},
         year = "2018",
        month = "Jun",
       volume = {859},
          eid = {L20},
        pages = {L20},
          doi = {10.3847/2041-8213/aab429},
archivePrefix = {arXiv},
       eprint = {1803.03265},
 primaryClass = {astro-ph.HE},
       adsurl = {https://ui.adsabs.harvard.edu/\#abs/2018ApJ...859L..20D}
}

@ARTICLE{Dai21review,
       author = {{Dai}, Jane Lixin and {Lodato}, Giuseppe and {Cheng}, Roseanne},
        title = "{The Physics of Accretion Discs, Winds and Jets in Tidal Disruption Events}",
      journal = {\ssr},
         year = 2021,
        month = feb,
       volume = {217},
       number = {1},
          eid = {12},
        pages = {12},
          doi = {10.1007/s11214-020-00747-x},
       adsurl = {https://ui.adsabs.harvard.edu/abs/2021SSRv..217...12D}
}

@ARTICLE{Evans89,
   author = {{Evans}, C.~R. and {Kochanek}, C.~S.},
    title = "{The tidal disruption of a star by a massive black hole}",
  journal = {\apjl},
     year = 1989,
    month = nov,
   volume = 346,
    pages = {L13-L16},
      doi = {10.1086/185567},
   adsurl = {http://adsabs.harvard.edu/abs/1989ApJ...346L..13E}
}

@ARTICLE{Gezari12,
   author = {{Gezari}, S. and {Chornock}, R. and {Rest}, A. and {Huber}, M.~E. and 
	{Forster}, K. and {Berger}, E. and {Challis}, P.~J. and {Neill}, J.~D. and 
	{Martin}, D.~C. and {Heckman}, T. and {Lawrence}, A. and {Norman}, C. and 
	{Narayan}, G. and {Foley}, R.~J. and {Marion}, G.~H. and {Scolnic}, D. and 
	{Chomiuk}, L. and {Soderberg}, A. and {Smith}, K. and {Kirshner}, R.~P. and 
	{Riess}, A.~G. and {Smartt}, S.~J. and {Stubbs}, C.~W. and {Tonry}, J.~L. and 
	{Wood-Vasey}, W.~M. and {Burgett}, W.~S. and {Chambers}, K.~C. and 
	{Grav}, T. and {Heasley}, J.~N. and {Kaiser}, N. and {Kudritzki}, R.-P. and 
	{Magnier}, E.~A. and {Morgan}, J.~S. and {Price}, P.~A.},
    title = "{An ultraviolet-optical flare from the tidal disruption of a helium-rich stellar core}",
  journal = {\nat},
archivePrefix = "arXiv",
   eprint = {1205.0252},
 primaryClass = "astro-ph.CO",
     year = 2012,
    month = may,
   volume = 485,
    pages = {217-220},
      doi = {10.1038/nature10990},
   adsurl = {http://adsabs.harvard.edu/abs/2012Natur.485..217G}
}

@ARTICLE{Gezari21review,
       author = {{Gezari}, Suvi},
        title = "{Tidal Disruption Events}",
      journal = {\araa},
         year = 2021,
        month = sep,
       volume = {59},
          doi = {10.1146/annurev-astro-111720-030029},
archivePrefix = {arXiv},
       eprint = {2104.14580},
 primaryClass = {astro-ph.HE},
       adsurl = {https://ui.adsabs.harvard.edu/abs/2021ARA&A..59...21G}
}

@ARTICLE{Guillochon13,
   author = {{Guillochon}, J. and {Ramirez-Ruiz}, E.},
    title = "{Hydrodynamical Simulations to Determine the Feeding Rate of Black Holes by the Tidal Disruption of Stars: The Importance of the Impact Parameter and Stellar Structure}",
  journal = {\apj},
archivePrefix = "arXiv",
   eprint = {1206.2350},
 primaryClass = "astro-ph.HE",
     year = 2013,
    month = apr,
   volume = 767,
      eid = {25},
    pages = {25},
      doi = {10.1088/0004-637X/767/1/25},
   adsurl = {http://adsabs.harvard.edu/abs/2013ApJ...767...25G}
}

@ARTICLE{Holoien16_14li,
   author = {{Holoien}, T.~W.-S. and {Kochanek}, C.~S. and {Prieto}, J.~L. and 
	{Stanek}, K.~Z. and {Dong}, S. and {Shappee}, B.~J. and {Grupe}, D. and 
	{Brown}, J.~S. and {Basu}, U. and {Beacom}, J.~F. and {Bersier}, D. and 
	{Brimacombe}, J. and {Danilet}, A.~B. and {Falco}, E. and {Guo}, Z. and 
	{Jose}, J. and {Herczeg}, G.~J. and {Long}, F. and {Pojmanski}, G. and 
	{Simonian}, G.~V. and {Szczygie{\l}}, D.~M. and {Thompson}, T.~A. and 
	{Thorstensen}, J.~R. and {Wagner}, R.~M. and {Wo{\'z}niak}, P.~R.
	},
    title = "{Six months of multiwavelength follow-up of the tidal disruption candidate ASASSN-14li and implied TDE rates from ASAS-SN}",
  journal = {\mnras},
archivePrefix = "arXiv",
   eprint = {1507.01598},
 primaryClass = "astro-ph.HE",
     year = 2016,
    month = jan,
   volume = 455,
    pages = {2918-2935},
      doi = {10.1093/mnras/stv2486},
   adsurl = {http://adsabs.harvard.edu/abs/2016MNRAS.455.2918H}
}

@ARTICLE{Hung19,
       author = {{Hung}, T. and {Cenko}, S.~B. and {Roth}, Nathaniel and {Gezari}, S. and {Veilleux}, S. and {van Velzen}, Sjoert and {Gaskell}, C. Martin and {Foley}, Ryan J. and {Blagorodnova}, N. and {Yan}, Lin and {Graham}, M.~J. and {Brown}, J.~S. and {Siebert}, M.~R. and {Frederick}, Sara and {Ward}, Charlotte and {Gatkine}, Pradip and {Gal-Yam}, Avishay and {Yang}, Yi and {Schulze}, S. and {Dimitriadis}, G. and {Kupfer}, Thomas and {Shupe}, David L. and {Rusholme}, Ben and {Masci}, Frank J. and {Riddle}, Reed and {Soumagnac}, Maayane T. and {van Roestel}, J. and {Dekany}, Richard},
        title = "{Discovery of Highly Blueshifted Broad Balmer and Metastable Helium Absorption Lines in a Tidal Disruption Event}",
      journal = {\apj},
         year = 2019,
        month = jul,
       volume = {879},
       number = {2},
          eid = {119},
        pages = {119},
          doi = {10.3847/1538-4357/ab24de},
archivePrefix = {arXiv},
       eprint = {1903.05637},
 primaryClass = {astro-ph.HE},
       adsurl = {https://ui.adsabs.harvard.edu/abs/2019ApJ...879..119H}
}

@ARTICLE{Kara16,
   author = {{Kara}, E. and {Miller}, J.~M. and {Reynolds}, C. and {Dai}, L.
	},
    title = "{Relativistic reverberation in the accretion flow of a tidal disruption event}",
  journal = {\nat},
archivePrefix = "arXiv",
   eprint = {1606.06736},
 primaryClass = "astro-ph.HE",
     year = 2016,
    month = jul,
   volume = 535,
    pages = {388-390},
      doi = {10.1038/nature18007},
   adsurl = {http://adsabs.harvard.edu/abs/2016Natur.535..388K}
}

@ARTICLE{Kasen06,
   author = {{Kasen}, D. and {Thomas}, R.~C. and {Nugent}, P.},
    title = "{Time-dependent Monte Carlo Radiative Transfer Calculations for Three-dimensional Supernova Spectra, Light Curves, and Polarization}",
  journal = {\apj},
   eprint = {astro-ph/0606111},
     year = 2006,
    month = nov,
   volume = 651,
    pages = {366-380},
      doi = {10.1086/506190},
   adsurl = {http://adsabs.harvard.edu/abs/2006ApJ...651..366K}
}

@ARTICLE{Lodato11,
       author = {{Lodato}, Giuseppe and {Rossi}, Elena M.},
        title = "{Multiband light curves of tidal disruption events}",
      journal = {\mnras},
         year = 2011,
        month = jan,
       volume = {410},
       number = {1},
        pages = {359-367},
          doi = {10.1111/j.1365-2966.2010.17448.x},
archivePrefix = {arXiv},
       eprint = {1008.4589},
 primaryClass = {astro-ph.CO},
       adsurl = {https://ui.adsabs.harvard.edu/abs/2011MNRAS.410..359L}
}

@ARTICLE{Loeb97,
   author = {{Loeb}, A. and {Ulmer}, A.},
    title = "{Optical Appearance of the Debris of a Star Disrupted by a Massive Black Hole}",
  journal = {\apj},
   eprint = {astro-ph/9703079},
     year = 1997,
    month = nov,
   volume = 489,
    pages = {573-578},
      doi = {10.1086/304814},
   adsurl = {http://adsabs.harvard.edu/abs/1997ApJ...489..573L}
}

@ARTICLE{Garcia2018,
       author = {{Gallegos-Garcia}, Monica and {Law-Smith}, Jamie and {Ramirez-Ruiz}, Enrico},
        title = "{Tidal Disruptions of Main-sequence Stars of Varying Mass and Age: Inferences from the Composition of the Fallback Material}",
      journal = {\apj},
         year = 2018,
        month = apr,
       volume = {857},
       number = {2},
          eid = {109},
        pages = {109},
          doi = {10.3847/1538-4357/aab5b8},
archivePrefix = {arXiv},
       eprint = {1801.03497},
 primaryClass = {astro-ph.HE},
       adsurl = {https://ui.adsabs.harvard.edu/abs/2018ApJ...857..109G}
}

@ARTICLE{Lu2020,
       author = {{Lu}, Wenbin and {Bonnerot}, Cl{\'e}ment},
        title = "{Self-intersection of the fallback stream in tidal disruption events}",
      journal = {\mnras},
         year = 2020,
        month = feb,
       volume = {492},
       number = {1},
        pages = {686-707},
          doi = {10.1093/mnras/stz3405},
archivePrefix = {arXiv},
       eprint = {1904.12018},
 primaryClass = {astro-ph.HE},
       adsurl = {https://ui.adsabs.harvard.edu/abs/2020MNRAS.492..686L}
}

@ARTICLE{Metzger2022,
       author = {{Metzger}, Brian D.},
        title = "{Cooling Envelope Model for Tidal Disruption Events}",
      journal = {\apjl},
         year = 2022,
        month = sep,
       volume = {937},
       number = {1},
          eid = {L12},
        pages = {L12},
          doi = {10.3847/2041-8213/ac90ba},
archivePrefix = {arXiv},
       eprint = {2207.07136},
 primaryClass = {astro-ph.HE},
       adsurl = {https://ui.adsabs.harvard.edu/abs/2022ApJ...937L..12M}
}

@ARTICLE{Parkinson20,
       author = {{Parkinson}, Edward J. and {Knigge}, Christian and {Long}, Knox S. and {Matthews}, James H. and {Higginbottom}, Nick and {Sim}, Stuart A. and {Hewitt}, Henrietta A.},
        title = "{Accretion disc winds in tidal disruption events: ultraviolet spectral lines as orientation indicators}",
      journal = {\mnras},
         year = 2020,
        month = jun,
       volume = {494},
       number = {4},
        pages = {4914-4929},
          doi = {10.1093/mnras/staa1060},
archivePrefix = {arXiv},
       eprint = {2004.07727},
 primaryClass = {astro-ph.HE},
       adsurl = {https://ui.adsabs.harvard.edu/abs/2020MNRAS.494.4914P}
}

@ARTICLE{Parkinson22,
       author = {{Parkinson}, Edward J. and {Knigge}, Christian and {Matthews}, James H. and {Long}, Knox S. and {Higginbottom}, Nick and {Sim}, Stuart A. and {Mangham}, Samuel W.},
        title = "{Optical line spectra of tidal disruption events from reprocessing in optically thick outflows}",
      journal = {\mnras},
         year = 2022,
        month = mar,
       volume = {510},
       number = {4},
        pages = {5426-5443},
          doi = {10.1093/mnras/stac027},
archivePrefix = {arXiv},
       eprint = {2201.01535},
 primaryClass = {astro-ph.HE},
       adsurl = {https://ui.adsabs.harvard.edu/abs/2022MNRAS.510.5426P}
}

@ARTICLE{Piran15,
   author = {{Piran}, T. and {Svirski}, G. and {Krolik}, J. and {Cheng}, R.~M. and 
	{Shiokawa}, H.},
    title = "{Disk Formation Versus Disk Accretion--What Powers Tidal Disruption Events?}",
  journal = {\apj},
archivePrefix = "arXiv",
   eprint = {1502.05792},
 primaryClass = "astro-ph.HE",
     year = 2015,
    month = jun,
   volume = 806,
      eid = {164},
    pages = {164},
      doi = {10.1088/0004-637X/806/2/164},
   adsurl = {http://adsabs.harvard.edu/abs/2015ApJ...806..164P}
}

@ARTICLE{Wu2018,
       author = {{Wu}, Samantha and {Coughlin}, Eric R. and {Nixon}, Chris},
        title = "{Super-Eddington accretion in tidal disruption events: the impactof realistic fallback rates on accretion rates}",
      journal = {\mnras},
         year = 2018,
        month = aug,
       volume = {478},
       number = {3},
        pages = {3016-3024},
          doi = {10.1093/mnras/sty971},
archivePrefix = {arXiv},
       eprint = {1804.06410},
 primaryClass = {astro-ph.HE},
       adsurl = {https://ui.adsabs.harvard.edu/abs/2018MNRAS.478.3016W}
}

@ARTICLE{Yuan12,
       author = {{Yuan}, Feng and {Wu}, Maochun and {Bu}, Defu},
        title = "{Numerical Simulation of Hot Accretion Flows. I. A Large Radial Dynamical Range and the Density Profile of Accretion Flow}",
      journal = {\apj},
         year = 2012,
        month = dec,
       volume = {761},
       number = {2},
          eid = {129},
        pages = {129},
          doi = {10.1088/0004-637X/761/2/129},
archivePrefix = {arXiv},
       eprint = {1206.4157},
 primaryClass = {astro-ph.HE},
       adsurl = {https://ui.adsabs.harvard.edu/abs/2012ApJ...761..129Y}
}

@ARTICLE{Rees88,
   author = {{Rees}, M.~J.},
    title = "{Tidal disruption of stars by black holes of 10 to the 6th-10 to the 8th solar masses in nearby galaxies}",
  journal = {\nat},
     year = 1988,
    month = jun,
   volume = 333,
    pages = {523-528},
      doi = {10.1038/333523a0},
   adsurl = {http://adsabs.harvard.edu/abs/1988Natur.333..523R}
}

@ARTICLE{Rossi21review,
       author = {{Rossi}, E.~M. and {Stone}, N.~C. and {Law-Smith}, J.~A.~P. and {Macleod}, M. and {Lodato}, G. and {Dai}, J.~L. and {Mandel}, I.},
        title = "{The Process of Stellar Tidal Disruption by Supermassive Black Holes}",
      journal = {\ssr},
         year = 2021,
        month = apr,
       volume = {217},
       number = {3},
          eid = {40},
        pages = {40},
          doi = {10.1007/s11214-021-00818-7},
       adsurl = {https://ui.adsabs.harvard.edu/abs/2021SSRv..217...40R}
}

@ARTICLE{Bowen1934,
       author = {{Bowen}, I.~S.},
        title = "{The Excitation of the Permitted O III Nebular Lines}",
      journal = {\pasp},
         year = 1934,
        month = jun,
       volume = {46},
       number = {271},
        pages = {146},
          doi = {10.1086/124435},
       adsurl = {https://ui.adsabs.harvard.edu/abs/1934PASP...46..146B}
}

@ARTICLE{Roth16,
   author = {{Roth}, N. and {Kasen}, D. and {Guillochon}, J. and {Ramirez-Ruiz}, E.
	},
    title = "{The X-Ray through Optical Fluxes and Line Strengths of Tidal Disruption Events}",
  journal = {\apj},
archivePrefix = "arXiv",
   eprint = {1510.08454},
 primaryClass = "astro-ph.HE",
     year = 2016,
    month = aug,
   volume = 827,
      eid = {3},
    pages = {3},
      doi = {10.3847/0004-637X/827/1/3},
   adsurl = {http://adsabs.harvard.edu/abs/2016ApJ...827....3R}
}

@ARTICLE{Roth17,
       author = {{Roth}, Nathaniel and {Kasen}, Daniel},
        title = "{What Sets the Line Profiles in Tidal Disruption Events?}",
      journal = {\apj},
         year = 2018,
        month = mar,
       volume = {855},
       number = {1},
          eid = {54},
        pages = {54},
          doi = {10.3847/1538-4357/aaaec6},
archivePrefix = {arXiv},
       eprint = {1707.02993},
 primaryClass = {astro-ph.HE},
       adsurl = {https://ui.adsabs.harvard.edu/abs/2018ApJ...855...54R}
}

@ARTICLE{Roth20review,
       author = {{Roth}, Nathaniel and {Rossi}, Elena Maria and {Krolik}, Julian and {Piran}, Tsvi and {Mockler}, Brenna and {Kasen}, Daniel},
        title = "{Radiative Emission Mechanisms}",
      journal = {\ssr},
         year = 2020,
        month = oct,
       volume = {216},
       number = {7},
          eid = {114},
        pages = {114},
          doi = {10.1007/s11214-020-00735-1},
archivePrefix = {arXiv},
       eprint = {2008.01117},
 primaryClass = {astro-ph.HE},
       adsurl = {https://ui.adsabs.harvard.edu/abs/2020SSRv..216..114R}
}

@ARTICLE{Saxton21review,
       author = {{Saxton}, R. and {Komossa}, S. and {Auchettl}, K. and {Jonker}, P.~G.},
        title = "{Correction to: X-Ray Properties of TDEs}",
      journal = {\ssr},
         year = 2021,
        month = feb,
       volume = {217},
       number = {1},
          eid = {18},
        pages = {18},
          doi = {10.1007/s11214-020-00759-7},
archivePrefix = {arXiv},
       eprint = {2103.15442},
 primaryClass = {astro-ph.HE},
       adsurl = {https://ui.adsabs.harvard.edu/abs/2021SSRv..217...18S}
}

@ARTICLE{Strubbe09,
   author = {{Strubbe}, L.~E. and {Quataert}, E.},
    title = "{Optical flares from the tidal disruption of stars by massive black holes}",
  journal = {\mnras},
archivePrefix = "arXiv",
   eprint = {0905.3735},
     year = 2009,
    month = dec,
   volume = 400,
    pages = {2070-2084},
      doi = {10.1111/j.1365-2966.2009.15599.x},
   adsurl = {http://adsabs.harvard.edu/abs/2009MNRAS.400.2070S}
}

@ARTICLE{Ulmer99,
   author = {{Ulmer}, A.},
    title = "{Flares from the Tidal Disruption of Stars by Massive Black Holes}",
  journal = {\apj},
     year = 1999,
    month = mar,
   volume = 514,
    pages = {180-187},
      doi = {10.1086/306909},
   adsurl = {http://adsabs.harvard.edu/abs/1999ApJ...514..180U}
}

@ARTICLE{vanVelzen20review,
       author = {{van Velzen}, Sjoert and {Holoien}, Thomas W. -S. and {Onori}, Francesca and {Hung}, Tiara and {Arcavi}, Iair},
        title = "{Optical-Ultraviolet Tidal Disruption Events}",
      journal = {\ssr},
         year = 2020,
        month = oct,
       volume = {216},
       number = {8},
          eid = {124},
        pages = {124},
          doi = {10.1007/s11214-020-00753-z},
archivePrefix = {arXiv},
       eprint = {2008.05461},
 primaryClass = {astro-ph.HE},
       adsurl = {https://ui.adsabs.harvard.edu/abs/2020SSRv..216..124V}
}

@ARTICLE{Svirski2017,
       author = {{Svirski}, Gilad and {Piran}, Tsvi and {Krolik}, Julian},
        title = "{Elliptical Accretion and Low Luminosity from High Accretion Rate Stellar Tidal Disruption Events}",
      journal = {\mnras},
         year = 2017,
        month = may,
       volume = {467},
       number = {2},
        pages = {1426-1432},
          doi = {10.1093/mnras/stx117},
archivePrefix = {arXiv},
       eprint = {1508.02389},
 primaryClass = {astro-ph.HE},
       adsurl = {https://ui.adsabs.harvard.edu/abs/2017MNRAS.467.1426S}
}

@ARTICLE{Bu2023,
       author = {{Bu}, De-Fu and {Qiao}, Erlin and {Yang}, Xiao-Hong},
        title = "{Radiative hydrodynamical simulations of super-Eddington accretion flow in tidal disruption event: the accretion flow and wind}",
      journal = {\mnras},
         year = 2023,
        month = aug,
       volume = {523},
       number = {3},
        pages = {4136-4145},
          doi = {10.1093/mnras/stad1696},
archivePrefix = {arXiv},
       eprint = {2306.04313},
 primaryClass = {astro-ph.HE},
       adsurl = {https://ui.adsabs.harvard.edu/abs/2023MNRAS.523.4136B}
}

@ARTICLE{Yang2024,
       author = {{Yang}, Hai and {Yuan}, Feng},
        title = "{Wind from the Hot Accretion Flow and Super-Eddington Accretion Flow}",
      journal = {arXiv e-prints},
         year = 2024,
        month = aug,
          eid = {arXiv:2408.16595},
        pages = {arXiv:2408.16595},
          doi = {10.48550/arXiv.2408.16595},
archivePrefix = {arXiv},
       eprint = {2408.16595},
 primaryClass = {astro-ph.HE},
       adsurl = {https://ui.adsabs.harvard.edu/abs/2024arXiv240816595Y}
}

@ARTICLE{Shlosman1993,
       author = {{Shlosman}, Isaac and {Vitello}, Peter},
        title = "{Winds from Accretion Disks: Ultraviolet Line Formation in Cataclysmic Variables}",
      journal = {\apj},
         year = 1993,
        month = may,
       volume = {409},
        pages = {372},
          doi = {10.1086/172670},
       adsurl = {https://ui.adsabs.harvard.edu/abs/1993ApJ...409..372S}
}

@ARTICLE{Holoien2014,
       author = {{Holoien}, T.~W. -S. and {Prieto}, J.~L. and {Bersier}, D. and {Kochanek}, C.~S. and {Stanek}, K.~Z. and {Shappee}, B.~J. and {Grupe}, D. and {Basu}, U. and {Beacom}, J.~F. and {Brimacombe}, J. and {Brown}, J.~S. and {Davis}, A.~B. and {Jencson}, J. and {Pojmanski}, G. and {Szczygie{\l}}, D.~M.},
        title = "{ASASSN-14ae: a tidal disruption event at 200 Mpc}",
      journal = {\mnras},
         year = 2014,
        month = dec,
       volume = {445},
       number = {3},
        pages = {3263-3277},
          doi = {10.1093/mnras/stu1922},
archivePrefix = {arXiv},
       eprint = {1405.1417},
 primaryClass = {astro-ph.GA},
       adsurl = {https://ui.adsabs.harvard.edu/abs/2014MNRAS.445.3263H}
}

@ARTICLE{Andreoni2022,
       author = {{Andreoni}, Igor and {Coughlin}, Michael W. and {Perley}, Daniel A. and {Yao}, Yuhan and {Lu}, Wenbin and {Cenko}, S. Bradley and {Kumar}, Harsh and {Anand}, Shreya and {Ho}, Anna Y.~Q. and {Kasliwal}, Mansi M. and {de Ugarte Postigo}, Antonio and {Sagu{\'e}s-Carracedo}, Ana and {Schulze}, Steve and {Kann}, D. Alexander and {Kulkarni}, S.~R. and {Sollerman}, Jesper and {Tanvir}, Nial and {Rest}, Armin and {Izzo}, Luca and {Somalwar}, Jean J. and {Kaplan}, David L. and {Ahumada}, Tom{\'a}s and {Anupama}, G.~C. and {Auchettl}, Katie and {Barway}, Sudhanshu and {Bellm}, Eric C. and {Bhalerao}, Varun and {Bloom}, Joshua S. and {Bremer}, Michael and {Bulla}, Mattia and {Burns}, Eric and {Campana}, Sergio and {Chandra}, Poonam and {Charalampopoulos}, Panos and {Cooke}, Jeff and {D'Elia}, Valerio and {Das}, Kaustav Kashyap and {Dobie}, Dougal and {Ag{\"u}{\'\i} Fern{\'a}ndez}, Jos{\'e} Feliciano and {Freeburn}, James and {Fremling}, Cristoffer and {Gezari}, Suvi and {Goode}, Simon and {Graham}, Matthew J. and {Hammerstein}, Erica and {Karambelkar}, Viraj R. and {Kilpatrick}, Charles D. and {Kool}, Erik C. and {Krips}, Melanie and {Laher}, Russ R. and {Leloudas}, Giorgos and {Levan}, Andrew and {Lundquist}, Michael J. and {Mahabal}, Ashish A. and {Medford}, Michael S. and {Miller}, M. Coleman and {M{\"o}ller}, Anais and {Mooley}, Kunal P. and {Nayana}, A.~J. and {Nir}, Guy and {Pang}, Peter T.~H. and {Paraskeva}, Emmy and {Perley}, Richard A. and {Petitpas}, Glen and {Pursiainen}, Miika and {Ravi}, Vikram and {Ridden-Harper}, Ryan and {Riddle}, Reed and {Rigault}, Mickael and {Rodriguez}, Antonio C. and {Rusholme}, Ben and {Sharma}, Yashvi and {Smith}, I.~A. and {Stein}, Robert D. and {Th{\"o}ne}, Christina and {Tohuvavohu}, Aaron and {Valdes}, Frank and {van Roestel}, Jan and {Vergani}, Susanna D. and {Wang}, Qinan and {Zhang}, Jielai},
        title = "{A very luminous jet from the disruption of a star by a massive black hole}",
      journal = {\nat},
         year = 2022,
        month = dec,
       volume = {612},
       number = {7940},
        pages = {430-434},
          doi = {10.1038/s41586-022-05465-8},
archivePrefix = {arXiv},
       eprint = {2211.16530},
 primaryClass = {astro-ph.HE},
       adsurl = {https://ui.adsabs.harvard.edu/abs/2022Natur.612..430A}
}

@ARTICLE{Velzen2021,
       author = {{van Velzen}, Sjoert and {Gezari}, Suvi and {Hammerstein}, Erica and {Roth}, Nathaniel and {Frederick}, Sara and {Ward}, Charlotte and {Hung}, Tiara and {Cenko}, S. Bradley and {Stein}, Robert and {Perley}, Daniel A. and {Taggart}, Kirsty and {Foley}, Ryan J. and {Sollerman}, Jesper and {Blagorodnova}, Nadejda and {Andreoni}, Igor and {Bellm}, Eric C. and {Brinnel}, Valery and {De}, Kishalay and {Dekany}, Richard and {Feeney}, Michael and {Fremling}, Christoffer and {Giomi}, Matteo and {Golkhou}, V. Zach and {Graham}, Matthew J. and {Ho}, Anna. Y.~Q. and {Kasliwal}, Mansi M. and {Kilpatrick}, Charles D. and {Kulkarni}, Shrinivas R. and {Kupfer}, Thomas and {Laher}, Russ R. and {Mahabal}, Ashish and {Masci}, Frank J. and {Miller}, Adam A. and {Nordin}, Jakob and {Riddle}, Reed and {Rusholme}, Ben and {van Santen}, Jakob and {Sharma}, Yashvi and {Shupe}, David L. and {Soumagnac}, Maayane T.},
        title = "{Seventeen Tidal Disruption Events from the First Half of ZTF Survey Observations: Entering a New Era of Population Studies}",
      journal = {\apj},
         year = 2021,
        month = feb,
       volume = {908},
       number = {1},
          eid = {4},
        pages = {4},
          doi = {10.3847/1538-4357/abc258},
archivePrefix = {arXiv},
       eprint = {2001.01409},
 primaryClass = {astro-ph.HE},
       adsurl = {https://ui.adsabs.harvard.edu/abs/2021ApJ...908....4V}
}

@ARTICLE{Popovic2006,
       author = {{Popovic}, L.~C.},
        title = "{The Broad Line Region of AGN: Kinematics and Physics}",
      journal = {Serbian Astronomical Journal},
         year = 2006,
        month = dec,
       volume = {173},
        pages = {1},
          doi = {10.2298/SAJ0673001P},
archivePrefix = {arXiv},
       eprint = {astro-ph/0612712},
 primaryClass = {astro-ph},
       adsurl = {https://ui.adsabs.harvard.edu/abs/2006SerAJ.173....1P}
}

@ARTICLE{Laor2006,
       author = {{Laor}, Ari},
        title = "{Evidence for Line Broadening by Electron Scattering in the Broad-Line Region of NGC 4395}",
      journal = {\apj},
         year = 2006,
        month = may,
       volume = {643},
       number = {1},
        pages = {112-119},
          doi = {10.1086/502798},
archivePrefix = {arXiv},
       eprint = {astro-ph/0601688},
 primaryClass = {astro-ph},
       adsurl = {https://ui.adsabs.harvard.edu/abs/2006ApJ...643..112L}
}

@ARTICLE{Kollatschny2013,
       author = {{Kollatschny}, W. and {Zetzl}, M.},
        title = "{The shape of broad-line profiles in active galactic nuclei}",
      journal = {\aap},
         year = 2013,
        month = jan,
       volume = {549},
          eid = {A100},
        pages = {A100},
          doi = {10.1051/0004-6361/201219411},
archivePrefix = {arXiv},
       eprint = {1211.3065},
 primaryClass = {astro-ph.CO},
       adsurl = {https://ui.adsabs.harvard.edu/abs/2013A&A...549A.100K}
}

@ARTICLE{Kouroumpatzakis2025,
       author = {{Kouroumpatzakis}, K. and {Svoboda}, J.},
        title = "{Gas excitation in galaxies and active galactic nuclei with He II{\ensuremath{\lambda}}4686 and X-ray emission}",
      journal = {\aap},
         year = 2025,
        month = apr,
       volume = {696},
          eid = {A133},
        pages = {A133},
          doi = {10.1051/0004-6361/202453192},
archivePrefix = {arXiv},
       eprint = {2503.03496},
 primaryClass = {astro-ph.GA},
       adsurl = {https://ui.adsabs.harvard.edu/abs/2025A&A...696A.133K}
}

@ARTICLE{Frederick2021,
       author = {{Frederick}, Sara and {Gezari}, Suvi and {Graham}, Matthew J. and {Sollerman}, Jesper and {van Velzen}, Sjoert and {Perley}, Daniel A. and {Stern}, Daniel and {Ward}, Charlotte and {Hammerstein}, Erica and {Hung}, Tiara and {Yan}, Lin and {Andreoni}, Igor and {Bellm}, Eric C. and {Duev}, Dmitry A. and {Kowalski}, Marek and {Mahabal}, Ashish A. and {Masci}, Frank J. and {Medford}, Michael and {Rusholme}, Ben and {Smith}, Roger and {Walters}, Richard},
        title = "{A Family Tree of Optical Transients from Narrow-line Seyfert 1 Galaxies}",
      journal = {\apj},
         year = 2021,
        month = oct,
       volume = {920},
       number = {1},
          eid = {56},
        pages = {56},
          doi = {10.3847/1538-4357/ac110f},
archivePrefix = {arXiv},
       eprint = {2010.08554},
 primaryClass = {astro-ph.HE},
       adsurl = {https://ui.adsabs.harvard.edu/abs/2021ApJ...920...56F}
}

@ARTICLE{Trakhtenbrot2019,
       author = {{Trakhtenbrot}, Benny and {Arcavi}, Iair and {Ricci}, Claudio and {Tacchella}, Sandro and {Stern}, Daniel and {Netzer}, Hagai and {Jonker}, Peter G. and {Horesh}, Assaf and {Mej{\'\i}a-Restrepo}, Juli{\'a}n Esteban and {Hosseinzadeh}, Griffin and {Hallefors}, Valentina and {Howell}, D. Andrew and {McCully}, Curtis and {Balokovi{\'c}}, Mislav and {Heida}, Marianne and {Kamraj}, Nikita and {Lansbury}, George Benjamin and {Wyrzykowski}, {\L}ukasz and {Gromadzki}, Mariusz and {Hamanowicz}, Aleksandra and {Cenko}, S. Bradley and {Sand}, David J. and {Hsiao}, Eric Y. and {Phillips}, Mark M. and {Diamond}, Tiara R. and {Kara}, Erin and {Gendreau}, Keith C. and {Arzoumanian}, Zaven and {Remillard}, Ron},
        title = "{A new class of flares from accreting supermassive black holes}",
      journal = {Nature Astronomy},
         year = 2019,
        month = jan,
       volume = {3},
        pages = {242-250},
          doi = {10.1038/s41550-018-0661-3},
archivePrefix = {arXiv},
       eprint = {1901.03731},
 primaryClass = {astro-ph.GA},
       adsurl = {https://ui.adsabs.harvard.edu/abs/2019NatAs...3..242T}
}

@ARTICLE{Shirazi2012,
       author = {{Shirazi}, Maryam and {Brinchmann}, Jarle},
        title = "{Strongly star forming galaxies in the local Universe with nebular He II{\ensuremath{\lambda}}4686 emission}",
      journal = {\mnras},
         year = 2012,
        month = apr,
       volume = {421},
       number = {2},
        pages = {1043-1063},
          doi = {10.1111/j.1365-2966.2012.20439.x},
archivePrefix = {arXiv},
       eprint = {1201.1290},
 primaryClass = {astro-ph.CO},
       adsurl = {https://ui.adsabs.harvard.edu/abs/2012MNRAS.421.1043S}
}

\appendix

\renewcommand{\thefigure}{A\arabic{figure}} 
\setcounter{figure}{0} 

\section{Resolving Emission Lines With {\tt SEDONA}} \label{sec:method:line}

For modeling emission lines, it is necessary to resolve the bound-bound transitions within the nLTE framework implemented in {\tt SEDONA}. During each iteration of the radiative transfer simulation, the radiation field and gas temperature are updated. These quantities are then used to determine the ionization and excitation states of the gas, which in turn set the frequency-dependent opacity and emissivity. In the nLTE regime, the radiation field influences transition rates between excitation states through both bound-bound and bound-free processes. Assuming statistical equilibrium, these transition rates are computed to estimate the level populations.

{\tt SEDONA} treats bound-bound processes using an indirect approach to reduce computational cost. Based on the opacity at the frequency of a photon packet, the code determines whether an interaction corresponds to electron scattering or effective scattering (i.e., absorption followed by re-emission). In the case of effective scattering, a new photon frequency is drawn from the local emissivity distribution. As a result, a photon packet initially associated with one transition (e.g., hydrogen) can be re-emitted at a different frequency (e.g., oxygen). When averaged over a large ensemble of photon packets, this method reproduces the results of more explicit line-transfer approaches while requiring significantly less computational time and memory.

Finally, to accurately simulate line processes, it is essential to resolve the intrinsic width of the line, which is determined by thermal velocity broadening:
\begin{equation} \label{eq:width}
    \Delta v_T=\frac{\Delta \nu}{\nu}c = \sqrt{\frac{2k_BT}{m_{\rm ions}} }=4 \times 10^6 {\rm \ {\rm \ cm \ s^{-1}}\ }\Big(\frac{T}{10^5K}\Big)^{0.5} \Big(\frac{A}{1}\Big)^{-0.5}
\end{equation}
where $\Delta v_T$ represents the thermal velocity broadening, $T$ is the temperature, $m_{\rm ions}$ is the mass of the atomic ion, and $A$ is the atomic number in question. Resolving this narrow intrinsic width requires several frequency bins across the line profile, which can be computationally expensive and memory-intensive.

To address this issue, we introduce an artificial intrinsic velocity broadening of $v_{\rm art, \ broad}=10^7 {\rm \ {\rm \ cm \ s^{-1}}}$. In our case, this value is smaller than the minimum bulk gas velocity ($3\times 10^7 {\rm \ {\rm \ cm \ s^{-1}}}$) and therefore does not affect the overall spectral properties. Instead, it improves numerical efficiency by accelerating effective scattering in optically thick regions.

The simulations presented here achieve a factor of four higher frequency resolution and employ an artificial broadening that is two orders of magnitude smaller than that used in \citet{Roth16}. This improvement allows for more accurate treatment of resonant scattering and, in particular, the Bowen fluorescence process. We have verified that our results are robust by testing smaller values of $v_{\rm art, \ broad}$, which yield consistent spectra.

\section{The Choice of the Atomic Library} \label{App:atom}
When performing Monte Carlo radiative transfer simulations, the choice of atomic data is critical. While a more comprehensive atomic library includes a larger number of transitions, it also substantially increases computational cost. For this reason, we adopt a simplified atomic library for the main simulations in this work, consisting of 1,982 energy levels and 13,024 line transitions. We also perform comparison runs using a more extensive library with 7,591 energy levels and 82,446 transitions.

\begin{figure}
    \centering
    \includegraphics[width=0.8\linewidth]{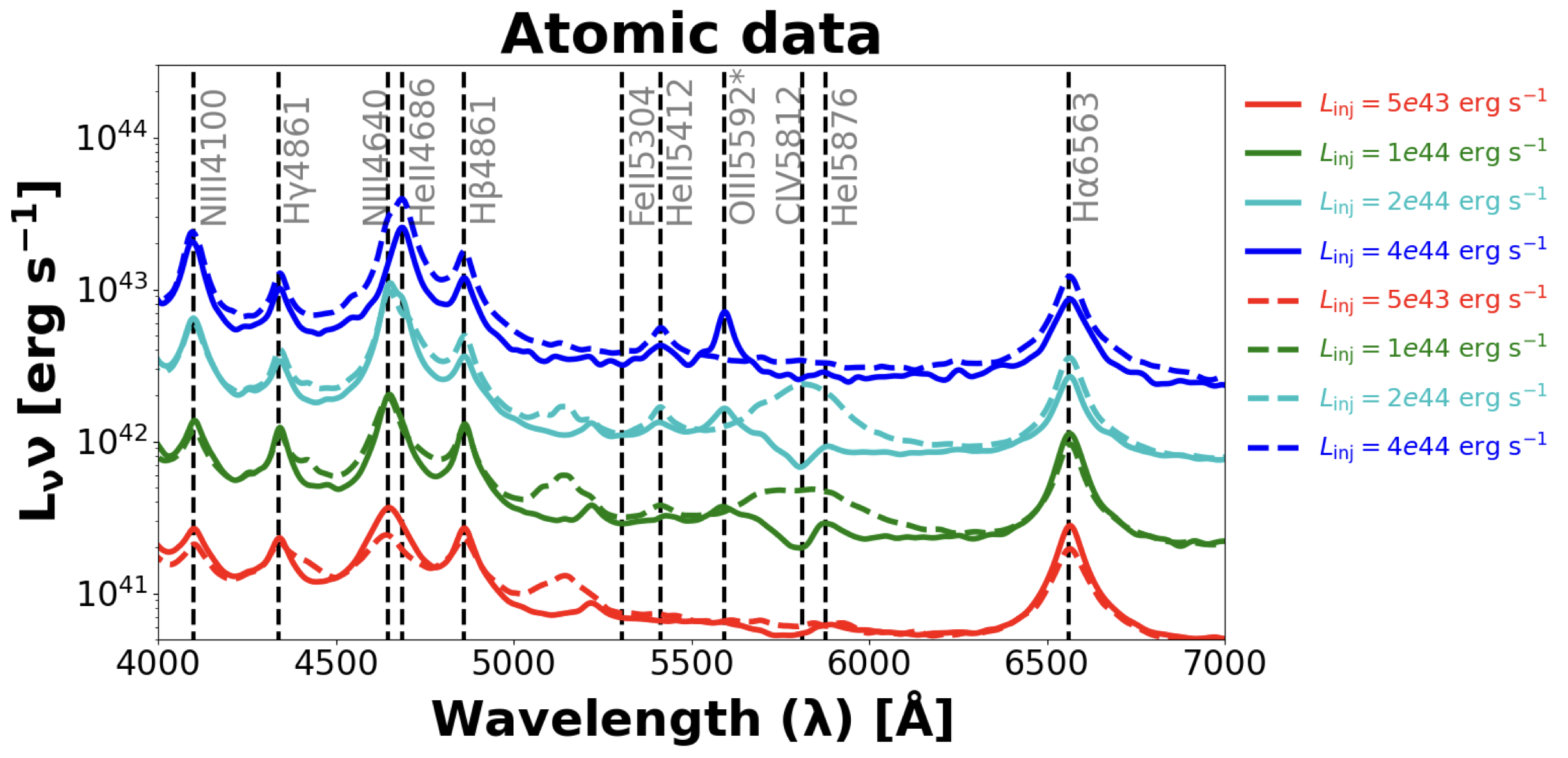}
    \caption{Comparison of spectra generated using different atomic libraries. The dashed curves correspond to simulations using the more extensive atomic dataset, while the solid curves show results from the simplified library adopted in this work. The detailed atomic model reproduces the overall spectral trends while eliminating artificial features, such as \oiiif, that arise in the simplified treatment.}
    \label{fig:atomic}
\end{figure}

As shown in Figure~\ref{fig:atomic}, both atomic libraries produce consistent results for the continuum and the primary emission features. However, differences arise in the treatment of weaker lines. In particular, the simplified atomic dataset artificially enhances certain transitions, such as \oiiif~(solid line). This enhancement is caused by the absence of recombination pathways from \oiv~to \oiii~in the reduced atomic model. These features disappear when using the more complete atomic dataset (dashed line), indicating that they are artifacts of the simplified treatment.

\section{Effect from the Compactness of the Envelope} \label{sec:appendix_He}
\begin{figure}
    \centering
    \includegraphics[width=0.8\linewidth]{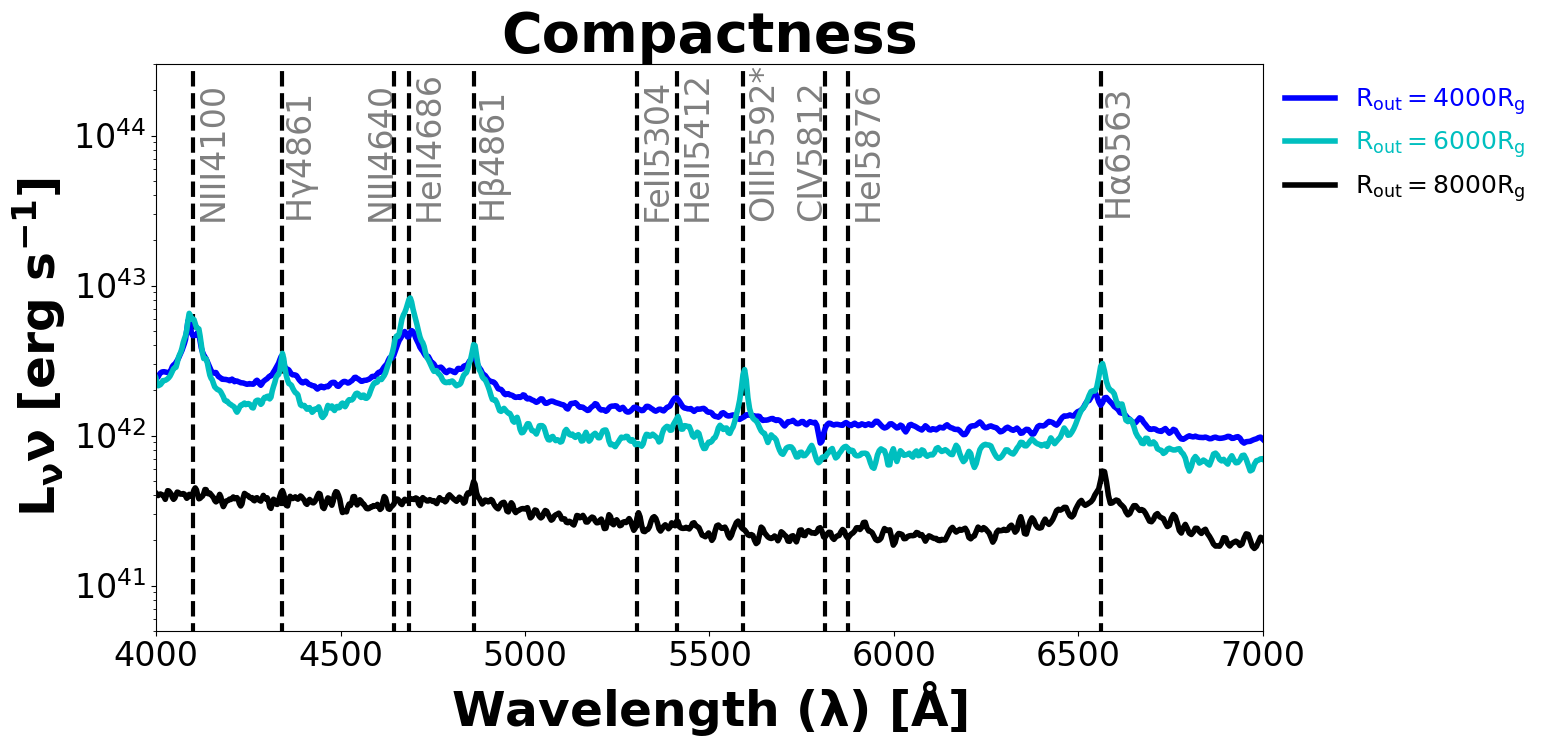}
    \caption{Effect of envelope compactness on the He-to-H line ratio. To isolate this effect, we fix the density slope and inner density corresponding to an envelope mass of $M = 0.1\,M_\odot$, while varying the outer radius at $4000\,R_g$, $6000\,R_g$ (fiducial), and $8000\,R_g$. The injected luminosity is held fixed at $L_{\rm inj} = 4 \times 10^{44}~\text{erg s}^{-1}$. As the envelope becomes more compact (smaller $R_{\rm out}$), the \heiia/\ha~line ratio increases, driven by the preferential suppression of Balmer emission, consistent with \citet{Roth16}.}
    \label{fig:compactness}
\end{figure}

Building on the work of \citet{Roth16}, we examine how the compactness of the reprocessing envelope influences the He-to-H line ratio. We find that reducing the outer radius $R_{\rm out}$ produces a more compact photosphere, which significantly enhances the \heiia/\ha~flux ratio, as shown in Figure~\ref{fig:compactness}. This behavior arises from the different formation depths of these lines: \heiia~originates closer to the inner, hotter regions of the envelope, whereas \ha~is produced predominantly at larger radii. As $R_{\rm out}$ decreases, the outer layers are truncated, preferentially suppressing \ha~emission while leaving \heiia~relatively unaffected, thereby increasing the He-to-H ratio.

A key physical parameter that can influence the effective compactness is the black hole mass $M_{\rm BH}$. For a fixed envelope mass and a structure defined in gravitational units, a larger $M_{\rm BH}$ spreads the debris over a greater physical volume. This geometric dilution reduces the mean density and optical depth of the envelope, leading to a higher overall ionization state. We therefore suggest that $M_{\rm BH}$ may act as a fundamental driver of photospheric compactness, imprinting itself on the observed spectroscopic properties and potentially contributing to the observed tendency for He-dominated TDEs to be associated with higher-mass black holes.

\end{document}